\documentstyle[preprint,aps,prc,rotate,array,dcolumn]{revtex}
\begin{document}
\input{psfig}
\input{epsf}
\def\Im{\mbox{\sl Im\ }}
\def\pd{\partial}
\def\oln{\overline}
\def\olft{\overleftarrow}
\def\ds{\displaystyle}
\def\bgreek#1{\mbox{\boldmath $#1$ \unboldmath}}
\def\sla#1{\slash \hspace{-2.5mm} #1}
\newcommand{\bra}{\langle}
\newcommand{\ket}{\rangle}
\newcommand{\vep}{\varepsilon}
\newcommand{\met}{{\mbox{\scriptsize met}}}
\newcommand{\lab}{{\mbox{\scriptsize lab}}}
\newcommand{\cm}{{\mbox{\scriptsize cm}}}
\newcommand{\mcal}{\mathcal}
\newcommand{\Del}{$\Delta$}
\newcommand{\g}{{\rm g}}
\long\def\Omit#1{}
\long\def\omit#1{\small #1}
\def\beq{\begin{equation}}
\def\eeq{\end{equation} }
\def\bea{\begin{eqnarray}}
\def\eea{\end{eqnarray}}
\def\eqref#1{Eq.~(\ref{eq:#1})}
\def\eqlab#1{\label{eq:#1}}
\def\figref#1{Fig.~\ref{fig:#1}}
\def\figlab#1{\label{fig:#1}}
\def\tabref#1{Table \ref{tab:#1}}
\def\tablab#1{\label{tab:#1}}
\def\secref#1{Section~\ref{sec:#1}}
\def\seclab#1{\label{sec:#1}}
\def\VYP#1#2#3{{\bf #1}, #3 (#2)}  
\def\NP#1#2#3{Nucl.~Phys.~\VYP{#1}{#2}{#3}}
\def\NPA#1#2#3{Nucl.~Phys.~A~\VYP{#1}{#2}{#3}}
\def\NPB#1#2#3{Nucl.~Phys.~B~\VYP{#1}{#2}{#3}}
\def\PL#1#2#3{Phys.~Lett.~\VYP{#1}{#2}{#3}}
\def\PLB#1#2#3{Phys.~Lett.~B~\VYP{#1}{#2}{#3}}
\def\PR#1#2#3{Phys.~Rev.~\VYP{#1}{#2}{#3}}
\def\PRC#1#2#3{Phys.~Rev.~C~\VYP{#1}{#2}{#3}}
\def\PRD#1#2#3{Phys.~Rev.~D~\VYP{#1}{#2}{#3}}
\def\PRL#1#2#3{Phys.~Rev.~Lett.~\VYP{#1}{#2}{#3}}
\def\FBS#1#2#3{Few-Body~Sys.~\VYP{#1}{#2}{#3}}
\def\AP#1#2#3{Ann.~of Phys.~\VYP{#1}{#2}{#3}}
\def\ZP#1#2#3{Z.\ Phys.\  \VYP{#1}{#2}{#3}}
\def\ZPA#1#2#3{Z.\ Phys.\ A\VYP{#1}{#2}{#3}}
\def\half{\mbox{\small{$\frac{1}{2}$}}}
\def\quarter{\mbox{\small{$\frac{1}{4}$}}}
\def\nn{\nonumber}
\newlength{\PicSize}
\newlength{\FormulaWidth}
\newlength{\DiagramWidth}
\newcommand{\vslash}[1]{#1 \hspace{-0.5 em} /}
\def\olaf{\marginpar{Mod-Olaf}}
\def\her{\marginpar{$\Longleftarrow$}}
\def\bel{\marginpar{$\Downarrow$}}
\def\abo{\marginpar{$\Uparrow$}}

\tighten


\title{Pion-nucleon amplitude near threshold: the sigma-term and
scattering lengths beyond few loops}

\author{S. Kondratyuk}
\address{
TRIUMF, 4004 Wesbrook Mall, Vancouver, British Columbia, Canada V6T 2A3}

\date{\today}

\maketitle

\begin{abstract}

The pion-nucleon amplitude is considered in the vicinity of 
the elastic scattering threshold within a relativistic dynamical model
dressing the $\pi N N$ and $\pi N \Delta$ vertices self-consistently 
with an infinite number of meson loops.
The dressing is formulated as solution of a system of coupled 
integral equations incorporating unitarity, crossing symmetry and analyticity
constraints. The calculated scattering lengths and the sigma-term
agree with recent data analyses.  
The dressing is important in this model both below and at threshold.
The contribution of the $\Delta$ resonance is discussed, 
including effects of the consistent dressing of the $\pi N\Delta$ vertex. 
A comparison with the approaches of
chiral perturbation theory and the Bethe-Salpeter equation is outlined.

\end{abstract}



\section{Introduction} \seclab{intro}

The pion-nucleon scattering amplitude near
the physical threshold is an interesting object to study for
a number of reasons. At the threshold point itself, the amplitude is
proportional to the s-wave scattering lengths, whose values
are known to be strongly constrained by chiral symmetry~\cite{Wei66}. 
For the amplitude below threshold, one can establish other chiral
low-energy theorems~\cite{Adl65,Che71} 
involving such quantities as the nucleon sigma-term~\cite{Rey74} and thus
related to the pattern of the 
explicit chiral symmetry breaking of QCD~\cite{Don86}. 
To extract the sigma-term from scattering data 
one usually analyses the amplitude at the Cheng-Dashen point. Although
corresponding to unphysical kinematics, this point is of special importance
because both pions are on-shell there and the difference
between the amplitude and the sigma-term is minimal~\cite{Che71,Gas88,Ols00}.

Chiral perturbation theory 
has been used to study loop corrections to the low-energy theorems and, 
in particular, to calculate the sigma-term~\cite{Gas81}. However, the
near-threshold region is significantly affected by the presence of
singularities which may make a non-relativistic perturbative expansion
unreliable~\cite{Bec01}.
In general, the pion-nucleon amplitude is ill-defined (and hence non-analytic)
at the Cheng-Dashen point and threshold is a branch point
dictated by unitarity.
To obey unitarity for pion-nucleon scattering exactly, one can
solve the relativistic Bethe-Salpeter equation with a tree-level potential; 
this yields a good description of the phase shifts 
and scattering lengths
as well as allows one to calculate the sigma-term~\cite{Lah99}. 
At the same time, the models based on the Bethe-Salpeter equation usually do 
not preserve crossing symmetry (however, see~\cite{Lut02}) which plays an
important role in the derivation of the low-energy theorems~\cite{Che71}.
  
In this paper the pion-nucleon 
amplitude is studied in the near-threshold region
using a relativistic dynamical model which incorporates
essential constraints from unitarity, analyticity and crossing symmetry. 
The effective lagrangian of the model includes pions, nucleons, 
the $\Delta$ resonance, the $\rho$ and $\sigma$ mesons. 
The main distinguishing feature of this approach is a special method of 
calculating pion-nucleon and
other meson-baryon loop corrections to free propagators and bare vertices. An
infinite series of loops is summed up by solving 
a system of coupled integral equations for the dressed vertices 
and propagators. These equations are formulated so that
constraints from unitarity, crossing symmetry and analyticity are fulfilled.
This formalism was expounded in Refs.~\cite{Kon99}, with a simplified 
treatment of the $\Delta$ resonance. In the present paper,
not only the $\pi N N$ vertex and the nucleon self-energy, but also
the $\pi N \Delta$ vertex and the $\Delta$ self-energy are dressed up to
infinite order.
The dressing procedure of~\cite{Kon99} was 
extended in~\cite{Knd00} to include photons while preserving gauge-invariance. 
Unitarity of the coupled-channel S-matrix above threshold was 
ensured since the dressing is consistent with the application of 
the dressed vertices and propagators in the K-matrix approach.  
In this way a good description of intermediate-energy 
pion-nucleon scattering, pion photoproduction and Compton
scattering was obtained, and the nucleon electromagnetic
polarisabilities were evaluated and found to be in agreement with
experiment~\cite{Kon01,Kon02}.
Since all parameters of the model (resonance coupling constants 
and a regularising cutoff) were fixed at the intermediate energies
in Ref.~\cite{Kon01}, the present calculation of the pion-nucleon 
amplitude below and at threshold is determined solely by the loop dynamics.

The outline of the paper is as follows.
The pion-nucleon amplitude at threshold and at the Cheng-Dashen point
is defined in \secref{pinampl}, where also relevant low-energy theorems are
cited. The formalism of the model is described in \secref{model}, including 
the integral equations for the dressed vertices and propagators and the method
of their solution.   
In \secref{expl_pin} the invariant pion-nucleon amplitudes in the near-threshold
region are explicitly expressed via the dressed functions. 
The main results of the paper will be presented in \secref{nearthresh_coef}. 
In particular, we will examine 
effects of multiple meson loops, of the 
$\Delta$ resonance and of the $\sigma$ and $\rho$ mesons, both below and at
threshold. The role of the $\Delta$ will be discussed further,
including the effects of the dressing of the 
$\pi N \Delta$ vertex. We will argue that 
the analyticity constraints incorporated in the dressing procedure 
are essential for the description of {\em both} the scattering lengths and 
the subthreshold coefficients in the {\em same} dynamical approach.  
Our formalism and results will be compared with
the approaches of chiral perturbations theory and the Bethe-Salpeter equation in
\secref{compar}. Concluding remarks are made in \secref{concl}.

\section{Pion-nucleon amplitude near threshold} \seclab{pinampl}

The standard isospin decomposition of the pion-nucleon amplitude is~\cite{Che57}
\beq
M_{\alpha \beta}=\delta_{\alpha \beta} M^+ + {1 \over 2}
[\tau_\alpha,\tau_\beta] M^- ,
\eqlab{amp_isosp}
\eeq
where $\tau_\alpha$ are Pauli matrices for the pion isospins. 
The spin structure of the amplitude is
\beq
M^{\pm} = \overline{u}(p') \left\{ A^{\pm} + 
\frac{\vslash{k}^{\prime}+\vslash{k}}{2}B^{\pm} \right\} u(p)
=
\overline{u}(p') \left\{ D^{\pm} - {1\over{4 m}}
[\vslash{k}',\vslash{k}] B^{\pm} \right\} u(p),
\eqlab{amp_sp}
\eeq
where $k$ and $k'$ ($p$ and $p'$) are the four-momenta of the initial and final 
pions (nucleons), respectively, $u(p)$ is the Dirac four-spinor.
The invariant amplitudes $A^{\pm}$, $B^{\pm}$, $D^{\pm}$ depend on
the Mandelstam variables $s=(p+k)^2$, $u=(p-k')^2$ and $t=(k-k')^2$.
We will also use the standard kinematic variables
$\nu=(s-u)/(4m)$ and $\nu_B=(t-2 \mu^2)/(4m)$, where $m=0.939$ GeV and 
$\mu=0.138$ GeV are the nucleon and pion masses. The two sets of
invariant functions in \eqref{amp_sp} are related as $D=A+\nu B$.

In the following we shall calculate the
value of the pion-nucleon amplitude at the subthreshold
Cheng-Dashen point, i.e.~at
$\nu=0$, $t=2 \mu^2$, with both pions on shell, $k^2=k^{\prime 2}=\mu^2$. 
One quantity of special interest is the sigma-term, which is related to the
pion-nucleon amplitude at the Cheng-Dashen point as
\beq
\Sigma = F_\pi^2 \lim_{\nu \rightarrow 0} \overline{D}^+(\nu,t=2 \mu^2),
\eqlab{sigterm}
\eeq
where $F_\pi=92.4$ MeV is the pion decay constant.
The bar indicates that the 
tree-level amplitude evaluated with the pseudovector $\pi N N$ vertex 
(usually called the ``pseudovector Born contribution") is subtracted,
\begin{eqnarray}
\overline{B}^{+}&=&B^{+} - \frac{g^2}{m}
\frac{\nu}{\nu_B^2-\nu^2} \,, 
\eqlab{bplbar} \\
\overline{B}^{-}&=&B^{-} - \frac{g^2}{m} 
\left( \frac{\nu_B}{\nu_B^2-\nu^2}
- \frac{1}{2 m} \right)  \,,
\eqlab{bmibar} \\
\overline{D}^{+}&=&D^{+} - \frac{g^2}{m} 
\frac{\nu_B^2}{\nu_B^2-\nu^2} \,, 
\eqlab{dplbar} \\
\overline{D}^{-}&=&D^{-} - \frac{g^2}{m}
\left( \frac{\nu \, \nu_B}{\nu_B^2-\nu^2}
- \frac{\nu}{2 m} \right). 
\eqlab{dmibar}
\end{eqnarray}
We take the value $g=13.02$~\cite{Kor98} 
for the $\pi N N$ coupling constant. The physical masses and coupling constants
of the particles used in this calculation are summarised in \tabref{masscoupl}.
According to a chiral low-energy theorem~\cite{Che71}, 
the sigma-term \eqref{sigterm} at the Cheng-Dashen point equals the 
scalar form factor of the nucleon up to corrections of order 
${\mathcal{O}}(\mu^4)$.
The scalar form factor can be related to  
the explicit chiral symmetry breaking (see, e.g., 
Refs.~\cite{Gas81,Bec01}).
Another low-energy theorem~\cite{Adl65,Che71} concerns the amplitude $D^-$,
requiring that the coefficient
\beq
C = 2 F_\pi^2 \, \lim_{\nu \rightarrow 0}  
\frac{\overline{D}^-(\nu,t = 2 \mu^2)}{\nu}
\eqlab{coefc}
\eeq
should approach unity in the chiral limit (i.e.~for a 
vanishing pion mass), up to corrections of order
${\mathcal{O}}(\mu^2)$.

In addition to the pion-nucleon amplitude in the subthreshold region,
we shall also calculate the s-wave scattering lengths which characterise the
amplitude at the threshold point
$Th \equiv \left\{ s=(m+\mu)^2, u=(m-\mu)^2, t=0 \right\}$:
\beq
a^{1/2}=\left. \frac{D^+ + 2 D^-}{4 \pi (1+\mu/m)} \; \right|_{Th} \,,
\eqlab{a1}
\eeq
\beq
a^{3/2}=\left. \frac{D^+ - D^-}{4 \pi (1+\mu/m)} \; \right|_{Th} \,,
\eqlab{a3}
\eeq
corresponding to the total isospins $1/2$ and $3/2$, respectively. The
low-energy theorem~\cite{Wei66} asserts that at lowest order the numerators
in Eqs.~(\ref{eq:a1}) and (\ref{eq:a3}) equal $\mu/(F_\pi^2)$ and
$-\mu/(2 F_\pi^2)$, respectively. 

\section{Description of the model} \seclab{model}

Our calculation of the near-threshold pion-nucleon amplitude is based on
the approach of Refs.~\cite{Kon99,Knd00,Kon01}. However, 
the treatment of the $\Delta$ resonance is significantly 
improved in the present version of the model, as will be explained in more
detail below. In this section we describe our approach, 
focusing on the ingredients which are most 
relevant in the near-threshold region.  
 
\subsection*{Structure of the amplitude} \seclab{amp_struct}

The $\pi N$ amplitude below and at threshold is purely real. It this model 
it is constructed as
the sum of the s-, u-channel nucleon and $\Delta$ exchange graphs, plus
the t-channel $\rho$ and $\sigma$ meson exchange graphs,
\beq
M = M_s+M_u+M^\Delta_s+M^\Delta_u+M^{\rho \sigma}_t,
\eqlab{amp_sut}
\eeq
as shown in \figref{pin_pnn_pnd}.
These are not simple tree diagrams, but rather skeleton diagrams as they
comprise {\em dressed} vertices and propagators.
Being a solution of a system of coupled integral equations, 
the nucleon and $\Delta$ propagators and vertices are dressed with meson
loops up to infinite order, while
the $\rho$ and $\sigma$ propagators are
calculated in a one $\pi \pi$ loop approximation.
Thus the central element of the approach is 
the calculation of the dressed vertices and propagators, which will be
described in the following.

\subsection*{Structure of the dressed vertices and propagators
of the nucleon and $\Delta$} 
\seclab{vertprop_struct}

The $\pi N N$ vertex required throughout the dressing procedure
has only one of the nucleons off the mass shell with the other nucleon and the
pion being on-shell (the so-called half-off-shell vertex).
For an incoming off-shell nucleon with the four-momentum squared $p^2$,
the most general Lorentz- and CPT-covariant structure of such 
a vertex is~\cite{Kaz59}\footnote{We
use the fully relativistic formalism, with 
the metric tensor, $\gamma$ matrices 
and other general conventions of \cite{Bjo64}.}
\beq
\tau_\alpha \Gamma(p)= \tau_{\alpha}\,\gamma^5 \Big[ G_{PS}(p^2)+
\frac{\vslash{p}+m}{2 m} G_{PV}(p^2)\Big],
\eqlab{pinvert}
\eeq 
where $G_{PS,PV}(p^2)$ are pseudovector and pseudoscalar form factors,
to be computed below.
The Lorentz-invariant expression for the nucleon self-energy is
written in terms of two invariant functions $A(p^2)$ and $B(p^2)$:
\beq
\Sigma(p) = \Sigma_L(p) -(Z_2-1)(\vslash{p}-m)-Z_2\,\delta m \; , \;\;
\Sigma_L(p)=A(p^2)\vslash{p}+B(p^2)m \, ,
\eqlab{nucself}
\eeq
where $\Sigma_L$ denotes the loop contributions 
to the self-energy. The complete self-energy $\Sigma(p)$ contains also 
the counter-term contribution with renormalisation constants 
$Z_2$ and $\delta m$ adjusted to provide the correct pole 
properties \eqref{prop_renorm} of the dressed nucleon propagator~\cite{Wei96} 
\beq
S(p)=\frac{1}{\vslash{p}-m-\Sigma(p)}=
\frac{\vslash{p}+\xi(p^2)}{\alpha(p^2)[p^2-\xi^2(p^2)]},
\eqlab{nucprop}
\eeq
where for later use we have introduced the self-energy functions 
\beq
\alpha(p^2)=Z_2-A(p^2),\;\;\;\; 
\xi(p^2)=\frac{m B(p^2)+Z_2(m-\delta m)}{\alpha(p^2)}.
\eqlab{sefun}
\eeq
The nucleon self-energy 
will be computed consistently with the $\pi N N$ vertex.

We choose the following form of the $\pi N \Delta$ vertex: 
\beq
T_\alpha \,V_{\mu}(k,p)=
T_\alpha\,\frac{\vslash{p}k_{\mu} - 
(p \cdot k) \gamma_{\mu}}{m_\Delta^2} \,
F_{\pi N \Delta}((p-k)^2)\, G_\Delta(p^2),
\eqlab{delvert}
\eeq 
where $p$ and $k$ are the 4-momenta of an incoming $\Delta$ and of an outgoing
pion, respectively, and $m_\Delta$ is the mass of the $\Delta$.
The part $G_\Delta(p^2)$ of the form factor in \eqref{delvert} will
be calculated in the dressing procedure. The real function 
$F_{\pi N \Delta}((p-k)^2)$ depending on the nucleon momentum
is needed for convergence of the procedure.
The isospin transition operators $T_\alpha$ are defined by the 
relations~\cite{Eri88}
\beq
T_\alpha\,T_\beta^\dagger = \delta_{\alpha \beta}-
\frac{\tau_{\alpha}\tau_{\beta}}{3},\;\;\;\;\;\;
T_\alpha^\dagger\,T_\alpha = 1.
\eqlab{isostran}
\eeq
To keep the calculations tractable, the
$\pi N \Delta$ vertex in \eqref{delvert} is not chosen in the most general
Lorentz-covariant form (for comparison, throughout the calculations 
we maintain the most general structure of the $\pi N N$ 
vertex). It is important however that
the vertex \eqref{delvert} has the property of 
``gauge-invariance"~\cite{Pas98}, 
\beq
p \cdot V(k,p)=0,
\eqlab{gauge_inv}
\eeq 
which allows us to eliminate the background 
spin $1/2$ component of the $\Delta$ propagator~\cite{Rar41} and to
keep only the spin $3/2$ component
\beq
S_\Delta^{\mu \nu}(p)=
\frac{1}{\vslash{p} - m_\Delta - \Sigma_\Delta(p) } 
{\mathcal{P}}_{3/2}^{\mu \nu}(p)=
\frac{\vslash{p}+\omega(p^2)}
{\eta(p^2)[p^2-\omega^2(p^2)] } \, {\mathcal{P}}_{3/2}^{\mu \nu}(p),
\eqlab{delprop}
\eeq
where the spin 3/2 projection operator 
\beq
{\mathcal{P}}_{3/2}^{\mu \nu}(p) = g^{\mu \nu}-\frac{\gamma^\mu \gamma^\nu}{3}
-\frac{\vslash{p} \gamma^\mu p^\nu + p^\mu \gamma^\nu \vslash{p}}{3 p^2}.
\eqlab{proj32}
\eeq
Formulae completely analogous to Eqs.~(\ref{eq:nucself},\ref{eq:sefun})
hold for the $\Delta$ self-energy. Although treating the $\Delta$ as a pure
spin $3/2$ state does not improve the description of pion-nucleon scattering
phase shifts as compared to the conventional treatment~\cite{Lah99}, 
it significantly simplifies the dynamical calculation of the $\Delta$
self-energy: we need to compute only two self-energy functions
$A_\Delta(p^2)$ and $B_\Delta(p^2)$, instead of 10 
invariant functions~\cite{Kor97} 
which would be required if the spin $1/2$ background were not eliminated.  
  
\subsection*{Integral equations for the dressing and their solution} 
\seclab{dres}

The $\pi N N$ form factors $G_{PV,PS}(p^2)$, the nucleon self-energy functions
$A(p^2), B(p^2)$, the $\pi N \Delta$ form factor $G_\Delta(p^2)$ and the
$\Delta$ self-energy functions  $A_\Delta(p^2), B_\Delta(p^2)$ are
calculated by solving a system of coupled integral equations.
This amounts to dressing these two- and three-point Green's functions with
meson loops up to infinite order.
In the earlier version of the model~\cite{Kon99,Knd00,Kon01}
the $\Delta$ resonance was not treated 
completely consistently with the nucleon: 
the $\Delta$ self-energy was computed up to one
$\pi N$ loop only and the dressing of the $\pi N \Delta$ vertex was not
included. However, considering nucleon 
Compton scattering, we showed~\cite{Kon02} that such simplified
$\Delta$ dressing, while being generally adequate, 
can lead to problems at low energies.
Therefore, in the present work we refine
the dressing procedure so that the nucleon and $\Delta$ 
are now treated on the same footing.

The dressing equations will be formulated using the following notation.
A generic Green's function ${\mathcal{G}}(q)$ 
is a sum of independent Lorentz-structures 
(e.~g.~$1$, $\vslash{q}$, $\gamma_\mu$, etc.), 
each of which is multiplied with a
Lorentz-invariant function depending on $q^2$ (such as form factors or
self-energy functions). If we use only imaginary or only real parts of the
invariant functions, the result will be denoted as
${\mathcal{G}}_I(q)$ or ${\mathcal{G}}_R(q)$, respectively.
If ${\mathcal{G}}(q)$ is calculated from a loop integral,
then according to Cutkosky rules~\cite{Cut60}
${\mathcal{G}}_I(q)$ is proportional to the discontinuity of
the integral across the unitary cut 
in the complex $q^2$ plane (due to pinching poles of the propagators in the integrand) 
and ${\mathcal{G}}_R(q)$ is the principal-value part of 
the integral.\footnote{If the theory obeys analyticity (causality) constraints,
the pole and principal-value parts of a loop must be  
related to each other through a dispersion integral~\cite{Bog59}.}

In our case, the principal-value and pole parts of the dressed 
$\pi N N$ vertex are denoted as $\Gamma_{R}(p)$ and $\Gamma_{I}(p)$, 
respectively. The expression for  $\Gamma_{R}(p)$ or $\Gamma_{I}(p)$ 
is obtained by using, respectively,  only the real or only the imaginary
parts of the form factors $G_{PV,PS}(p^2)$ in
the right-hand side of \eqref{pinvert}. 
The same applies to the $\pi N \Delta$ vertex \eqref{delvert}:
to obtain $(V_\mu)_R(k,p)$ or $(V_\mu)_I(k,p)$ we  use only  
$\mbox{Re} G_\Delta(p^2)$ or $\mbox{Im} G_\Delta(p^2)$, respectively.
Similarly, the pole part $\Sigma_I(p)$ 
of the nucleon self-energy~\eqref{nucself} contains only 
$\mbox{Im} A(p^2)$ and $\mbox{Im} B(p^2)$, 
and the principal-value part $\Sigma_R(p)$ only 
$\mbox{Re} A(p^2)$ and $\mbox{Re} B(p^2)$. 
The pion propagator $D(k)= \left[ k^2 - \mu^2 + i0 \right]^{-1}$ 
does not get dressed, therefore its imaginary part comes from 
the on-shell pions: $D_I(k)=\delta(k^2-\mu^2) \theta(k_0)$. 
In the same way, we retain only the dominant pole contribution
to the discontinuity of the nucleon propagator:
$S_I(p)=(\vslash{p}+m) \delta(p^2-m^2) \theta(p_0)$. 
The resonance propagators do not have poles on the physical
Riemann sheet, so their discontinuous parts come solely from their 
self-energies. For example, the discontinuity
of the dressed $\Delta$ propagator~\eqref{delprop}
is obtained by keeping only the imaginary parts of its invariant functions:
\beq
(S_\Delta^{\mu \nu})_I(p) = \left\{ \vslash{p} \,
\mbox{Im} \frac{1}{\eta(p^2)[\, p^2-\omega^2(p^2) \,]} +
\mbox{Im} \frac{\omega(p^2)}{\eta(p^2)[\, p^2-\omega^2(p^2) \,]} \right\}
\theta(p_0) \,,
\eqlab{delprdisc}
\eeq 
and analogously for the $\rho$ and $\sigma$ mesons (unlike the 
$\Delta$, however, the propagators of the meson resonances
are dressed in a one $\pi \pi$ loop approximation only, as will be
discussed in more detail below and in Appendix A).


With the introduced notation, 
the system of dressing equations can be written 
\begin{eqnarray}
\Gamma_{I}(p) & = & {\ds \frac{1}{8 \pi^2}  \int d^4 k\, 
\Gamma_{R}(p'-k) S(p'-k) \overline{\Gamma}_{R}(p'-k) 
S_I(p-k) D_I(k) \Gamma_{R}(p) } \nn \\
& + & {\ds \frac{1}{6 \pi^2}  \int d^4 k\,
(V_\mu)_R(-k,p'-k) S_\Delta^{\mu \nu}(p'-k)
(\overline{V}_\nu)_R(-q,p'-k) } \nn \\
& & \hspace*{12mm} \times {\ds S_I(p-k) D_I(k) \Gamma_{R}(p) } \nn \\
& - & {\ds \frac{1}{6 \pi^2}  \int d^4 k\,
\Gamma_{R}(p'-k) S(p'-k) (V_\mu)_R(q,p-k) 
({S}_\Delta^{\mu \nu})_I(p-k) } \nn \\
& & \hspace*{12mm} \times D_I(k) (\overline{V}_\nu)_R(k,p-k) \,+\, 
\Gamma_{I}^{\rho \sigma}(p)\;,                              
\eqlab{dr_eq_n1} \\ \nn \\
\mbox{Re} \left\{ {{\ds G_{PV}} \atop {\ds G_{PS}}} \right\}(p^2)
&=& \left\{ {{\ds G^0_{PV}} \atop {\ds G^0_{PS}}} \right\}(p^2)
+ {\ds \frac{\mathcal{P}}{\pi}
\int_{(m+\mu)^2}^{\infty} \!\! dp^{\prime 2}\,
\frac{\mbox{Im} \left\{ {{\ds G_{PV}} \atop {\ds G_{PS}}} \right\} (p^{\prime 2})}
{p^{\prime 2}-p^2}\;, } 
\eqlab{dr_eq_n2} \\ \nn \\
\Sigma_I(p) &=& {\ds -\frac{3}{8 \pi^2}\overline{\Gamma}_{R}(p)
\int\!d^4 k\, S_I(p-k) D_I(k) \Gamma_{R}(p)} \;, 
\eqlab{dr_eq_n3} \\ \nn \\
\mbox{Re} \left\{ {{\ds A} \atop {\ds B}} \right\}(p^2) 
&=& {\ds \frac{\mathcal{P}}{\pi} \int_{(m+\mu)^2}^{\infty} \!\!
dp^{\prime 2}\, 
\frac{\mbox{Im} \left\{ {{\ds A} \atop {\ds B}} \right\} (p^{\prime 2})}
{p^{\prime 2}-p^2}\;, } 
\eqlab{dr_eq_n4} \\ \nn \\
(V_\mu)_I(q,p) & = & {\ds \frac{1}{4 \pi^2}  \int d^4 k\, 
\Gamma_{R}(p^{\prime}-k) S(p^{\prime}-k) 
\overline{\Gamma}_{R}(p^{\prime}-k) S_I(p-k) } \nn \\ 
& & \hspace*{12mm} \times D_I(k) (V_\mu)_R(k,p)  \nn \\
& + & {\ds \frac{1}{24 \pi^2}  \int d^4 k\,
(V_\nu)_R(-k,p'-k) S_\Delta^{\nu \lambda}(p'-k)   
(\overline{V}_\lambda)_R(-q,p'-k) } \nn \\
& & \hspace*{12mm} \times S_I(p-k) D_I(k) (V_\mu)_R(k,p) 
\, +\, (V^{\rho \sigma}_\mu)_I(q,p) \;, 
\eqlab{dr_eq_d1} \\ \nn \\
\mbox{Re} G_\Delta(p^2) &=& G_\Delta^0(p^2)+ {\ds \frac{\mathcal{P}}{\pi}
\int_{(m+\mu)^2}^{\infty} \!\! dp^{\prime 2}\,
\frac{\mbox{Im} G_\Delta(p^{\prime 2})}{p^{\prime 2}-p^2}\;, }  
\eqlab{dr_eq_d2} \\ \nn \\
\Sigma^{\Delta}_I(p) &=& {\ds 
\frac{ {\mathcal{P}}_{3/2}^{\nu \mu}(p) }{16 \pi^2}
\int\!d^4 k\,  
 (\overline{V}_{\mu})_R(k,p) S_I(p-k) D_I(k) (V_{\nu})_R(k,p) } \;, 
\eqlab{dr_eq_d3} \\ \nn \\   
\mbox{Re} \left\{ {{\ds A_{\Delta}} \atop {\ds B_{\Delta}}} \right\}(p^2) 
&=& {\ds \frac{\mathcal{P}}{\pi} \int_{(m+\mu)^2}^{\infty} \!\!
dp^{\prime 2}\, \frac{\mbox{Im} \left\{ 
{{\ds A_{\Delta}} \atop {\ds B_{\Delta}}} \right\} (p^{\prime 2})}
{p^{\prime 2}-p^2}\;, }  
\eqlab{dr_eq_d4}   
\end{eqnarray} 
where ${\mathcal{P}}$ denotes taking the 
principal-value of an integral, and isospin factors have been
absorbed in the coefficients on the right-hand side.  
The inhomogeneities $\Gamma_{I}^{\rho \sigma}(p)$ in \eqref{dr_eq_n1} and
$(V^{\rho \sigma}_\mu)_I(q,p)$ in \eqref{dr_eq_d1} contain  
$\rho$ and $\sigma$ mesons, as described in
Appendix A.
The real functions $G_{PV,PS}^0(p^2)$ and $G_\Delta^0$ are bare 
$\pi N N$ and $\pi N \Delta$ form factors, respectively.
To see that the system of dressing 
equations is neither under- nor over-determined, note that
Eqs.~(\ref{eq:dr_eq_n1},\ref{eq:dr_eq_n3}) and (\ref{eq:dr_eq_d3}) have
two independent spinor structures each. Thus  
Eqs.~(\ref{eq:dr_eq_n1}--\ref{eq:dr_eq_d4}) are 14 scalar 
equations for 14 scalar functions  
$\mbox{Im} G_{PV}(p^2)$, $\mbox{Im} G_{PS}(p^2)$, 
$\mbox{Im} A(p^2)$, $\mbox{Im} B(p^2)$, 
$\mbox{Im} G_\Delta(p^2)$, $\mbox{Im} A_\Delta(p^2)$, $\mbox{Im} B_\Delta(p^2)$,
$\mbox{Re} G_{PV}(p^2)$, $\mbox{Re} G_{PV}(p^2)$,
$\mbox{Re} A(p^2)$, $\mbox{Re} B(p^2)$, 
$\mbox{Re} G_\Delta(p^2)$, $\mbox{Re} A_\Delta(p^2)$, $\mbox{Re} B_\Delta(p^2)$.

Formally, Eqs.~(\ref{eq:dr_eq_n1}--\ref{eq:dr_eq_d4}) constitute a 
coupled system of nonlinear integral equations.
Despite its quite complicated analytic form, 
this system of equations has a rather transparent meaning 
(see \figref{eq_pnn_pnd}). The equations are solved by
iteration, starting with input bare form factors $G^0_{PV,PS}$, $G_\Delta^0$.
In the course of iteration one effectively sums up an infinite series 
of meson-loop corrections to the bare vertices and free propagators.
At each iteration step we first calculate the discontinuities of the
loop integrals through the Cutkosky rules; these pole parts are then
used in  dispersion integrals to compute the corresponding
principal-value parts. 
The details of the computation technique can be found in \cite{Kon99,Knd00}.
Here we will recapitulate only the most important points and
discuss the new issues arising due to the consistent incorporation of the
$\Delta$ resonance in the dressing procedure.

The use of bare form factors
$G_{PV,PS}^0(p^2)$ and $G_\Delta^0$
is necessary to regularise 
the equations.\footnote{An attempt to get rid of the bare 
form factors by using subtracted dispersion integrals in  
Eqs.~(\ref{eq:dr_eq_n2},\ref{eq:dr_eq_n4},\ref{eq:dr_eq_d2},\ref{eq:dr_eq_d4})
fails since each new iteration step would require more subtractions 
than the previous.}
We choose the purely pseudovector structure for the bare $\pi N N$ vertex, 
i.~e.~$G_{PS}^0(p^2)=0$, since 
the derivative coupling of pions at low energies is
dictated by chiral symmetry. The bare $\pi N N$ from factor is chosen in the
form 
\beq
G_{PV}^0(p^2)=f\,
\exp{\left[-\ln{2}\frac{(p^2-m^2)^2}{\Lambda_N^4}\right]},
\eqlab{bareff}
\eeq 
where $\Lambda_N^2$ is a half-width. 
The bare coupling constant $f \equiv f_{\pi N N}$ 
is adjusted so that the dressed $\pi N N$ vertex
is normalised on-shell to the physical $\pi N N$ coupling constant:
\beq
\lim_{{p \hspace{-0.35 em} /} \rightarrow m} \Gamma(p) = g \, \gamma^5 \,,
\eqlab{vert_renorm}
\eeq
where the ``sandwich" between the spinors of the initial and final nucleons 
is implicit. The usual renormalisation~\cite{Wei96} 
of the dressed nucleon propagator
\beq
\lim_{{p \hspace{-0.35 em} /} \rightarrow m} S(p) = \frac{1}
{{p \hspace{-0.45 em} /} - m}
\eqlab{prop_renorm}
\eeq
is imposed by adjusting the field renormalisation constant $Z_2 \equiv Z_2^N$ 
and the mass shift $\delta m \equiv \delta m_N$ (see 
Eqs.~({\ref{eq:nucself}) and (\ref{eq:nucprop})). 
The renormalisation of the $\pi N \Delta$ vertex and of the $\Delta$ propagator
is done similarly, except that now the pole properties are required of a
propagator with only the real parts of the self-energy
functions~\cite{Cut60,Kon99}. There are
three corresponding renormalisation constants: $f_{\pi N \Delta}$,
$Z_2^\Delta$ and $\delta m_\Delta$. Note
that due to the coupled nature of the dressing  
Eqs.~(\ref{eq:dr_eq_n1}--\ref{eq:dr_eq_d4})
the six renormalisation conditions for the vertices and propagators
can be obeyed only simultaneously. Thus $f_{\pi N N}$, $Z_2^N$,
$\delta m_N$, and $f_{\pi N \Delta}$, $Z_2^\Delta$,
$\delta m_\Delta$ are interdependent. 
The complete set of renormalisation constants obtained in the calculation 
are given in \tabref{renorm}. 

We stress that
the half-width $\Lambda_N^2$ is not a completely independent parameter: the
iteration procedure converges only for $\Lambda_N^2 < (\Lambda_N^{max})^2$, 
and it is important that $(\Lambda_N^{max})^2$ is much larger than the
energy scale due to the explicitly included particles. With the set of
parameters used in this calculation, $(\Lambda_N^{max})^2 \approx 3$ GeV$^2$. 
When a convergent solution does exist, it is reached in practice after
about 30 iteration steps for $\Lambda_N^2 \approx (\Lambda_N^{max})^2$.
The bare $\pi N \Delta$ form factor
$G_\Delta^0(p_\Delta^2)$ 
as well as the form factors $F_{\pi N \Delta}(p_N^2)$,
$F_{\rho \pi \pi}(p_{\rho}^2)$, $F_{\sigma \pi \pi}(p_{\sigma}^2)$, 
$F_{\rho N N}(p_N^2)$ and $F_{\sigma N N}(p_N^2)$, appearing in the
vertices with the resonances, 
have the same exponential form as \eqref{bareff}, but
peak at the masses of the corresponding particles (see Appendix A).
We assume the width $\Lambda_R^2$ of these bare form factors to be the same
for all resonances, and set it close 
to the maximal value allowed by the convergence
requirement. This is done in keeping with the general
emphasis of our approach on the loop dynamics as determined by the dressing
rather than on fitting additional parameters. Ideally, values of 
$\Lambda_R^2$ for different meson resonances should come from a dynamical 
dressing of these mesons on the same footing with the nucleon and the $\Delta$.
Such an extension of the model is certainly feasible,
although has not been done yet.

The effects of the loop corrections on the $\pi N N$ vertex are similar to
those discussed in detail in \cite{Kon99,Kon01}, where the $\Delta$ was not
dressed consistently. We mention the main points here. 
The dressing generates an energy-dependent admixture of
the pseudoscalar coupling, which at low energies remains much
smaller than the pseudovector component and becomes more prominent only at 
intermediate energies. 
It is important that the dressing does {\em not} allow for 
a large pseudoscalar admixture to develop in the low-energy region. 
The pseudovector form factor in narrowed by the dressing. 
This softening persists independently of the functional form of 
the bare form factor (provided the latter falls sufficiently 
fast at infinity) and is stronger for wider bare form factors. 

Like the half-widths $\Lambda_N^2$ and $\Lambda_R^2$, 
the coupling constants 
$g_{\rho N N}$, $\kappa_\rho$,
$g_{\sigma N N}$, $g_{\sigma \pi \pi}$ and $f_{\sigma \pi \pi}$
of the $\rho$ and $\sigma$ mesons
are mutually constrained by the requirement that a 
convergent solution of the dressing equations should exist.
\figref{param_dep} shows the area of convergence in the space of the
coupling constants $g_{\sigma N N}$ and $g_{\rho N N}$, the other parameters
being fixed at their values given in \tabref{param}. 
The convergence area has a nontrivial shape 
for small $g_{\sigma N N}$ and large $g_{\rho N N}$
(or for small $g_{\rho N N}$ and large $g_{\sigma N N}$). For these values
there is less cancellation between loops with $\sigma$'s 
and those with $\rho$'s, which precludes convergence.
Generalising the illustration in \figref{param_dep}, 
convergent solutions of Eqs.~(\ref{eq:dr_eq_n1}--\ref{eq:dr_eq_d4})
can be found only in a certain subspace of the space spanned by
$\Lambda_N^2$, $\Lambda_R^2$, $g_{\rho N N}$, $\kappa_\rho$, 
$g_{\sigma N N}$, $g_{\sigma \pi \pi}$ and $f_{\sigma \pi \pi}$.    
The remaining freedom in the parameters
was removed by calculating the pion-nucleon phase shifts
using the dressed K-matrix approach~\cite{Kon01} and comparing them with 
data analyses~\cite{Arn95} at intermediate energies. 
The dot in \figref{param_dep} corresponds to the thus optimised
coupling constants $g_{\rho N N}$ and $g_{\sigma N N}$.
The full set of relevant parameters from~\cite{Kon01} 
are reproduced in \tabref{param}. We emphasise that in fixing these values
we did not use any results of data analyses for
the scattering lengths or for the subthreshold amplitudes. 
Therefore there are {\em no free parameters} 
in the present calculation of the near-threshold amplitudes.

\section{Scattering amplitude in terms of the dressed vertices and propagators} 
\seclab{expl_pin}

Having solved the dressing equations, we proceed to evaluate the
invariant amplitudes $A^{\pm}$ and $B^{\pm}$ defined in \eqref{amp_sp}.
First we write the s- and u-channel diagrams from \eqref{amp_sut} in terms of
the dressed $\pi N N$, $\pi N \Delta$ vertices 
Eqs.~(\ref{eq:pinvert},\ref{eq:delvert})
and dressed nucleon, $\Delta$ propagators
Eqs.~(\ref{eq:nucprop},\ref{eq:delprop}). Then we add 
the t-channel diagrams written in terms of
the relevant meson vertices and propagators 
Eqs.~(\ref{eq:rhopp},\ref{eq:rhoprop},\ref{eq:rhonn},\ref{eq:sigpp},\ref{eq:sigmaprop},\ref{eq:signn}). 
At the considered kinematics, only the real parts of the
invariant functions of the two- and three-point Green's functions
enter in the amplitude.
After doing some straightforward algebra,
the contribution of the nucleon, $\rho$ and $\sigma$ exchange diagrams 
to the invariant amplitudes can be written  
\begin{eqnarray}
A^{+} & = &
\frac{ {\displaystyle G_{PS}^2(s)[\xi(s)-m] + \frac{G_{PS}(s)G_{PV}(s)}{m}[s-m^2]+
\frac{G_{PV}^2(s)}{4 m^2}[m+\xi(s)][s-m^2] } }{\alpha(s)[s-\xi^2(s)]} \nn \\
 & + &
\frac{ {\displaystyle G_{PS}^2(u)[\xi(u)-m] + \frac{G_{PS}(u)G_{PV}(u)}{m}[u-m^2]+
\frac{G_{PV}^2(u)}{4 m^2}[m+\xi(u)][u-m^2] } }{\alpha(u)[u-\xi^2(s)]} \nn \\ 
& - & \frac{ {\displaystyle g_{\sigma N N} F_{\sigma \pi \pi}(t) 
\left[g_{\sigma \pi \pi} \mu - f_{\sigma \pi \pi} 
\frac{k^{\prime 2}+k^2-t}{2 \mu} \right] } }
{Z^\sigma \,  [t-\zeta^2(t)]},  \eqlab{apl} \\
A^{-} & = &
\frac{ {\displaystyle G_{PS}^2(s)[\xi(s)-m] + \frac{G_{PS}(s)G_{PV}(s)}{m}[s-m^2]+
\frac{G_{PV}^2(s)}{4 m^2}[m+\xi(s)][s-m^2] } }{\alpha(s)[s-\xi^2(s)]} \nn \\
 & - &
\frac{ {\displaystyle G_{PS}^2(u)[\xi(u)-m] + \frac{G_{PS}(u)G_{PV}(u)}{m}[u-m^2]+
\frac{G_{PV}^2(u)}{4 m^2}[m+\xi(u)][u-m^2] } }{\alpha(u)[u-\xi^2(s)]} \nn \\ 
& + & \frac{ {\displaystyle g_{\rho N N} g_{\rho \pi \pi} \kappa_\rho 
F_{\rho \pi \pi}(t) (u-s)} }
{2 m Z^\rho \,  [t-\lambda^2(t)]},  \eqlab{ami} \\
B^{+} & = &
-\frac{ {\displaystyle G_{PS}^2(s) + \frac{G_{PS}(s)G_{PV}(s)}{m}[m+\xi(s)] +
\frac{G_{PV}^2(s)}{4 m^2}[m^2+s+2m\xi(s)] } }{\alpha(s)[s-\xi^2(s)]} \nn \\
 & + &
\frac{ {\displaystyle G_{PS}^2(u) + \frac{G_{PS}(u)G_{PV}(u)}{m}[m+\xi(u)] +
\frac{G_{PV}^2(u)}{4 m^2}[m^2+u+2m\xi(u)] } }{\alpha(u)[u-\xi^2(u)]}, \eqlab{bpl}
\\
B^{-} & = &
-\frac{ {\displaystyle G_{PS}^2(s) + \frac{G_{PS}(s)G_{PV}(s)}{m}[m+\xi(s)] +
\frac{G_{PV}^2(s)}{4 m^2}[m^2+s+2m\xi(s)] } }{\alpha(s)[s-\xi^2(s)]} \nn \\
 & - &
\frac{ {\displaystyle G_{PS}^2(u) + \frac{G_{PS}(u)G_{PV}(u)}{m}[m+\xi(u)] +
\frac{G_{PV}^2(u)}{4 m^2}[m^2+u+2m\xi(u)] } }{\alpha(u)[u-\xi^2(u)]} \nn \\
 & + &
\frac{ {\displaystyle g_{\rho N N} g_{\rho \pi \pi} (1+\kappa_\rho) 
F_{\rho \pi \pi}(t) } }
{Z^\rho \,  [t-\lambda^2(t)]}.  \eqlab{bmi}
\end{eqnarray}

The $\Delta$ exchange diagrams from \eqref{amp_sut} are given by
(restoring the isospin indices) 
\begin{eqnarray}
(M_s^{\Delta})_{\alpha \beta} & = & \left( \delta_{\alpha \beta} - 
\frac{\tau_\alpha \tau_\beta}{3} \right)\,
\frac{G_\Delta^2(s)}{m_\Delta^4 \, \eta(s) [s-\omega^2(s)]} 
\overline{u}(p') \left[ (\vslash{p}'+\vslash{k}')k'_\mu -
(p'+k') \cdot k' \, \gamma_\mu \right] \nn \\
& \times &
\left[\vslash{p}+\vslash{k}+\omega(s)\right]
{\mathcal{P}}_{3/2}^{\mu \nu}(p+k) \left[ (\vslash{p}+\vslash{k})k_\nu -
(p+k) \cdot k \, \gamma_\nu \right] u(p) \,, \eqlab{dels} \\
(M_u^{\Delta})_{\alpha \beta} & = & \left( \delta_{\alpha \beta} - 
\frac{\tau_\beta \tau_\alpha}{3} \right)\,
\frac{G_\Delta^2(u)}{m_\Delta^4 \, \eta(u) [u-\omega^2(u)]} 
\overline{u}(p') \left[ -(\vslash{p}'-\vslash{k})k_\mu +
(p'-k) \cdot k \, \gamma_\mu \right] \nn \\
& \times &
\left[\vslash{p}-\vslash{k}'+\omega(u)\right]
{\mathcal{P}}_{3/2}^{\mu \nu}(p-k') \left[ -(\vslash{p}-\vslash{k}')k'_\nu +
(p-k') \cdot k' \, \gamma_\nu \right] u(p) \,. \eqlab{delu}
\end{eqnarray}
For brevity we will not give the explicit decomposition of 
Eqs.~(\ref{eq:dels},\ref{eq:delu}) in terms of the invariant amplitudes.

\subsection*{Properties of analyticity and crossing symmetry} \seclab{anal}

Due to the use of the cutting rules and dispersion integrals in
the formulation of Eqs.~(\ref{eq:dr_eq_n1}--\ref{eq:dr_eq_d4}), the
two- and three-point Green's functions obtained by solving these equations
possess the correct analyticity structure associated
with the nucleon and $\Delta$ exchanges and obey the two-body ($\pi N$) 
unitarity. 
Furthermore, the $\pi N$ amplitude obeys the crossing symmetry 
requirements that $A^{+}$, $A^{-}/(s-u)$, $B^{+}/(s-u)$ and $B^{-}$ 
be invariant under the replacement
$s \leftrightarrow u$. The crossing is respected due to
our using the dressed two- and three-point Green's functions in {\em both}
s- and u-type diagrams. At the same time, the t-channel 
analyticity structure is not fully reproduced because
the t-channel cuts are taken into account only through the $\pi \pi$ 
loops in the $\rho$ and $\sigma$ propagators but
the loop corrections to the vertices with the $\rho$ and
$\sigma$ mesons are not included. Also, the
four-point one-particle irreducible diagrams--such as the box 
graph--are not included in the dressing. 

The omitted dressed one-particle irreducible four-point
diagrams can be thought of as being of order ${\mathcal{O}}(a^2)$ in a 
certain formal expansion~\cite{Kon01}, where the 
parameter $a$ characterises the level of
analyticity violation in the model. 
In this expansion, the lowest order
${\mathcal{O}}(a^0)$ corresponds to an amplitude with no dressing, in which case
the violation of analyticity in maximal.
The next order ${\mathcal{O}}(a^1)$ corresponds to an amplitude in which
the one-particle reducible (with respect to the s-channel cuts) 
graphs contain the dressed propagators and vertices. 
Thus, at order ${\mathcal{O}}(a^1)$ 
analyticity is restored at the level of two- and three-point Green's 
functions, as is done in present dressing procedure. 
The higher orders in $a$ are described by induction in terms of reducibility of n-point Green's
functions constituting the amplitude.

The parameter $a$ can be defined as follows. 
If a scattering amplitude
$T(\omega)$ of a process can be represented at small energies $\omega$
as a power series
\beq
T(\omega)=c_0+c_1 \omega + c_2 \omega^2 + \ldots,
\eqlab{form_exp}
\eeq
then the coefficients $c_i$ could, in principle, 
be computed {\em using the same model} but in two different ways: 
\begin{enumerate}
\item
Low-energy ($LE$) evaluation:
compute $c_i(LE)$ by evaluating $T(\omega)$
directly at low energies; 
\item
Sum-rule ($SR$) evaluation:
compute $c_i(SR)$ by calculating appropriate total cross sections and
integrating them in the sum rules corresponding to each particular coefficient.
\end{enumerate}
These two ways of evaluation should give identical coefficients 
$c_i(LE)=c_i(SR)=c_i$ 
provided the analyticity of the model is exact. 
In practical calculations 
there will always be discrepancies between the two  
methods, which can be used to estimate the
violation of analyticity in the model. Therefore
one can quantify the parameter $a$ as 
\beq
a \sim | c_i(LE) - c_i(SR) |  \,, 
\eqlab{a_to_c}
\eeq
where the choice of a particular coefficient from the series \eqref{form_exp}
could be decided by additional considerations.
This idea was tested for nucleon Compton scattering in 
Ref.~\cite{Kon02}, where the coefficients $c_1$ and $c_{2,3}$ were 
related to the anomalous magnetic moment and to nucleon polarisabilities,
respectively. 
The proposed formal expansion in the parameter $a$
can offer a systematic way of improving analyticity properties 
of dynamical approaches applicable at low and intermediate energies.
 
The properties of analyticity and crossing symmetry are crucial for 
the model to provide a good description of the amplitude 
both below and at the physical threshold.\footnote{A similar conclusion was 
reached in the
framework of the relativistic baryon chiral perturbation theory~\cite{Bec01}, 
where it was pointed out that the standard low-energy expansion does 
not reproduce the correct analyticity structure 
in the vicinity of singularities.}
To study this in more detail, we shall focus below on the 
sigma-term, the Adler-Weisberger coefficient $C$, as defined by 
Eqs.~(\ref{eq:sigterm},\ref{eq:coefc}), and on the scattering lengths,
as defined by Eqs.~(\ref{eq:a1},\ref{eq:a3}). We will collectively 
call these quantities the ``near-threshold coefficients".

\section{Near-threshold coefficients} \seclab{nearthresh_coef}
\subsubsection*{The sigma-term and coefficient $C$ 
at the Cheng-Dashen point}  \seclab{chdpoint}

On expanding the amplitudes in Eqs.~(\ref{eq:apl}--\ref{eq:delu}) 
around the Cheng-Dashen
point $s=u=m^2, t=2 \mu^2$ (with $k^2=k^{\prime 2}=\mu^2$)
and using the definitions Eqs.~(\ref{eq:sigterm},\ref{eq:coefc}), 
we obtain explicit formulae for the sigma-term and for the
coefficient $C$ in terms of the dressed vertices and propagators:  
\begin{eqnarray}
\Sigma&=&-F_\pi^2 \left\{  
\frac{G_{PS}^2(m^2)}{m \, \alpha(m^2)} +
\frac{g_{\sigma N N} \, g_{\sigma \pi \pi} \, \mu F_{\sigma \pi \pi}(2 \mu^2)}
{Z^\sigma\,[\,2 \mu^2 - \zeta^2(2 \mu^2)\,]} \right\} + \Sigma_\Delta,
\eqlab{sigterm_calc} \\
C&=&2 F_\pi^2 \left\{ 
\frac{G_{PS}(m^2)}{\alpha(m^2)} 
\left[ \frac{g}{m^2}-4G_{PS}'(m^2) \right]-\frac{G_{PS}^2(m^2)}{2 \alpha^2(m^2)}
\left[\frac{1}{m^2} - 4 \alpha'(m^2)  \right] \right. \nn  \\
&+& \left. \frac{ g_{\rho N N} \, g_{\rho \pi \pi}
(1-\kappa_{\rho})F_{\rho \pi \pi}(2 \mu^2)}
{Z^\rho \, [ \, 2 \mu^2-\lambda^2(2 \mu^2) \,]}  \right\} + C_\Delta,
\eqlab{coefc_calc}
\end{eqnarray}
where $\Sigma_\Delta$ and $C_\Delta$ contain the effects of
the $\Delta$ exchange in the s- and u-channel diagrams
(see Appendix B). In Eqs.~(\ref{eq:sigterm_calc})
and (\ref{eq:coefc_calc}) we have made use of the relations
\beq
\xi(m^2)=m, \;\;\;\; \xi'(m^2) = \frac{\alpha(m^2)-1}
{2 \, m \, \alpha(m^2)},\;\;\;\;
G_{PS}(m^2)+G_{PV}(m^2)=g,
\eqlab{ren_expl}
\eeq
which follow from the
renormalisation conditions Eqs.~(\ref{eq:vert_renorm},\ref{eq:prop_renorm}).

Eqs.~(\ref{eq:sigterm_calc}) and (\ref{eq:coefc_calc})
show that the subthreshold parameters $\Sigma$ and $C$ are
sensitive to the values and derivatives of
the dressed $\pi N N$ form factor $G_{PS}(p^2)$ and of the nucleon self-energy 
function $\alpha(p^2)$ at the nucleon pole. 
Since the pseudoscalar form factor in the dressed $\pi N N$ vertex is very small 
in the vicinity of $m^2$, the nucleon contribution in
Eqs.~(\ref{eq:sigterm_calc},\ref{eq:coefc_calc}) is much smaller than that 
of the $\sigma$ and $\rho$ mesons.
The $\sigma$ exchange in the t-channel gives a
dominant numerical contribution to the sigma-term
due to the presence of the non-derivative component 
$\sim g_{\sigma \pi \pi} \mu$ in the $\sigma \pi \pi$ 
vertex \eqref{sigpp}.  
The $\rho$ meson plays a similar role for the coefficient $C$.
It should be pointed out, however, that dissection of 
Eqs.~(\ref{eq:sigterm_calc},\ref{eq:coefc_calc}) into
a ``nucleon contribution", ``$\sigma$ and $\rho$ meson contributions" 
and a ``$\Delta$ contribution" 
can only be regarded as formal here: as already mentioned, 
Eqs.~(\ref{eq:dr_eq_n1}--\ref{eq:dr_eq_d4}) not only determine the 
nucleon and $\Delta$ dressing, but also strongly constrain
the other parameters of the model. 
For example,
an arbitrary change of the values of the $\rho$ and $\sigma$ coupling constants 
would formally change the dominant contributions in 
Eqs.~(\ref{eq:sigterm_calc},\ref{eq:coefc_calc}), but with these altered
coupling constants the dressing procedure might not converge at all!

In what follows we will discuss various ingredients of the dressing by 
comparing the ``Dressed"  and ``Bare" calculations. 
The former contains the full dressing with the meson loops whereas 
in the latter the bare vertices and free propagators have been used.
Note that since the $\rho \pi \pi$, $\rho N N$, $\sigma \pi \pi$ and
$\sigma N N$ vertices do not get dressed in the model, 
in both calculations they are equipped with the bare form factors
$F_{\pi N \Delta}(p_N^2)$,
$F_{\rho \pi \pi}(p_{\rho}^2)$, $F_{\sigma \pi \pi}(p_{\sigma}^2)$, 
$F_{\rho N N}(p_N^2)$ and $F_{\sigma N N}(p_N^2)$, as defined in
Appendix A.
The obtained values of the near-threshold coefficients are summarised in
\tabref{thresh_mod}. Results of several data analyses are quoted in the
last row. Various ingredients of the fully
dressed calculation are given in the other rows and
will be discussed below in more detail.

\subsubsection*{Pion-nucleon scattering lengths} \seclab{lengths}

The scattering lengths are evaluated by
substituting the explicit expressions for the amplitudes 
Eqs.~(\ref{eq:apl}--\ref{eq:delu})
into Eqs.~(\ref{eq:a1}) and (\ref{eq:a3}).
The obtained values are listed in \tabref{thresh_mod} for the
different calculations considered. The large effect of the dressing on 
$a^{1/2}$, in comparison with the small effect on $a^{3/2}$,
is a reflection of the 
nucleon s-channel graph being influenced by the dressing more than the
u-channel graph. 

\subsection*{Dependence on the bare form factor} \seclab{bare_dep}

To see how our results depend on the choice of the bare form factor 
$G^0_{PV}$ in \eqref{bareff},  
we did a calculation in which the width of the 
bare form factor was set to $\Lambda_N^2=2.8$ GeV$^2$ (i.~e.~near the
upper limit $(\Lambda_N^2)^{max}$ dictated by the convergence)  
and all the other parameters were kept as given in \tabref{param}. 
As was shown in \cite{Kon99},
despite using a much wider form factor, such a calculation leads to
intermediate-energy phase shifts which are similar to 
those obtained in the basic calculation with $\Lambda_N^2=1.8$ GeV$^2$.
With this variation of the bare width, the
sigma-term and the coefficient $C$ change by less than 3\%.  
The sensitivity of the scattering lengths is similarly small.
The important point here is that even using quite different bare form factors 
the model yields results which are comparable with data analyses {\em both} at 
the Cheng-Dashen point and at threshold. 
As long as the iteration procedure converges, 
the loop dynamics depend only weakly on the details of the bare form factor. 
The usage of the exponential bare form factor is also not essential and 
dipole-like bare form factors lead to similar dressed vertices.
By contrast, the role of the dressing is quite significant.
This can be seen by comparing the rows labelled ``Dressed" and 
``Bare" in \tabref{thresh_mod}.

\subsection*{Effects of the dressing of the $\sigma$ and $\rho$ propagators} 
\seclab{rhosig_role}

In the fully dressed calculation, the self-energies of
the $\sigma$ and $\rho$ mesons are dressed with one $\pi \pi$ loop, 
as detailed in Appendix A.
By using the free propagators instead, we obtain the results listed in the
rows ``Free $\sigma$" and ``Free $\rho$" in \tabref{thresh_mod}. 
The $\rho$ exchange contributes solely
to the isospin-odd amplitudes while the $\sigma$ exchange acts in the 
isospin-even channel, which is
explicitly shown in Eqs.~(\ref{eq:apl}--\ref{eq:bmi}). 
Hence the sigma-term $\Sigma$ (being calculated from $D^+$ in
\eqref{sigterm}) is oblivious to the treatment of the $\rho$ meson while the 
coefficient $C$ (being calculated from $D^-$ in \eqref{coefc}) is not 
affected by the $\sigma$ meson.

\subsection*{Role of the $\Delta$ resonance} \seclab{delta}

\subsubsection*{$\Delta$ pole contribution}\seclab{del_pol}

By the $\Delta$ contribution to the pion-nucleon amplitudes 
one usually means the contribution of the
s- and u-channel $\Delta$ exchange diagrams. 
This definition may be used for a comparison
of our results with those of the chiral perturbation theory, 
where the $\Delta$ is usually not included as an explicit 
field in the lagrangian and thus does not appear in the
loops~\cite{Ber95,Fet98,Bec01}.\footnote{There exist approaches~\cite{Ber93}  
including the $\Delta$ explicitly in chiral 
lagrangians, in which case the $\Delta$ does enter in the loops.
Such extensions of the standard chiral perturbation theory will not be
discussed here.}
By comparing the rows ``Dressed" and ``No $\Delta$ poles" in 
\tabref{thresh_mod}, we see that the pole contribution of the $\Delta$ is 
small both below and at threshold (the
explicit formulae are given in Appendix B). 
For example, it changes the sigma-term by $0.26$ MeV, 
which is somewhat smaller than
the values typically found in other calculations~\cite{Ols00,Bec01,Ber96}. 
However, a
quantitative comparison of our results with other approaches should be
carried out carefully since typically one
retains a spin $1/2$ background in the $\Delta$ propagator
(see, e.~g., \cite{Ber95,Bec01}).   
By contrast, in this model we deal with the pure spin 3/2 $\Delta$ 
due to the gauge-invariant structure of the 
$\pi N \Delta$ vertex \eqref{delvert}.

\subsubsection*{Effects of the dressing of the $\pi N \Delta$ vertex}
\seclab{del_consdr}

In addition to the s- and u-channel exchanges,
the $\Delta$ resonance enters in the loop corrections to the
$\pi N N$ and $\pi N \Delta$ vertices {\em both} of which are dressed up to infinite order. 
Such a contribution of the $\Delta$ through dressing has not been
considered before in the context of the near-threshold $\pi N$ amplitude.
Due to the coupled nature of 
Eqs.~(\ref{eq:dr_eq_n1}--\ref{eq:dr_eq_d4}), 
the $\Delta$ dressing affects the $\pi N$ amplitude
directly as well as through its effects on the $\pi N N$ vertex and nucleon
propagator.
 
In the subthreshold region, the value of the 
sigma-term is slightly decreased by
the $\Delta$ dressing, as can be seen by comparing the rows
``Dressed" and ``Bare $\Delta$" in \tabref{thresh_mod}
(in the latter calculation the bare $\pi N \Delta$ vertex and the free
$\Delta$ propagator were used).
Comparison of these results with the column ``No $\Delta$ poles"
shows that the dressing moderates even further the small 
pole contribution of the $\Delta$ to the sigma-term. 
Notably, the effect of the consistent dressing of the $\pi N \Delta$ vertex is to 
decrease the sigma-term and to increase the coefficient $C$, whereas the
$\Delta$ exchange has the opposite effect. This suggests that meson
loops should be treated with special care in dynamical calculations of the
$\Delta$ contribution to the subthreshold $\pi N$ amplitude.

\subsection*{Effects of multiple loops} \seclab{multi}   
   
The two- and three-point functions obtained as a solution of 
Eqs.~(\ref{eq:dr_eq_n1}--\ref{eq:dr_eq_d4}) comprise an infinite series of
meson-loop corrections. The effect of multiple
meson loops on the near-threshold amplitude can be seen from
the row labelled ``One loop" in \tabref{thresh_mod}. This is a calculation in which
the $\pi N N$ and $\pi N \Delta$ vertices, as well as
the nucleon, $\Delta$, $\rho$ and $\sigma$ self-energies, 
include only one-loop corrections.
These vertices and self-energies are obtained by
iterating Eqs.~(\ref{eq:dr_eq_n1}--\ref{eq:dr_eq_d4}) only once
after discarding from them the effective two-loop contributions 
(the latter are the integrals
containing one of the cut resonance propagators $(S_\Delta^{\mu \nu})_I$,
$(D_\rho^{\mu \nu})_I$ or $(D_\sigma)_I$). 
To focus on the genuine effects of the loops, 
no parameters were readjusted in this calculation,
except the renormalisation constants
$f_{\pi N N},\, f_{\pi N \Delta},\, Z_2^N,\,
Z_2^\Delta, \delta m_N,\, \delta m_\Delta$. 
Note that in this case the $\pi N$ scattering amplitude 
is not exactly an amplitude including only 
one-loop corrections to the tree-level approximation. 
Nevertheless, taking into account that most of the
effects of the dressing are due to the vertices,  
the difference between this ``one loop" and
the fully dressed calculations can serve as a reasonable estimate 
of the effects beyond few perturbative loop corrections in the amplitude. 
\tabref{thresh_mod} shows that the multiple-loop effects are small, 
being comparable in magnitude with the contribution of the 
$\Delta$ resonance (Ref.~\cite{Ber96} gives an 
upper limit of $\approx 2$ MeV
for the $\Delta$ contribution to the sigma-term).  
The value of the sigma-term comes mostly from the one-loop calculation,
whereas the multiple loops and the $\Delta$ give 
contributions which are beyond the precision of the present data analyses. 
 
\Omit{
}
     
\section{Comparison with the approaches of
chiral perturbation theory and the Bethe-Salpeter equation}
\seclab{compar}

The near-threshold parameters obtained in this model 
are compared in \tabref{chpt_comp} with calculations in 
chiral perturbation theory ($\chi$PT) and with those based on the 
Bethe-Salpeter equation (BSE). We quote the results of
the heavy-baryon and relativistic (with the infrared regularisation) baryon  
formulations of chiral expansions, 
referred to as HB$\chi$PT and RB$\chi$PT, respectively.  
There is some similarity between the present model and the
other approaches, as well as a number of differences. 
In principle, it would be desirable to compare
the loops calculated in this model with the chiral loops and with the
loops generated by the BSE at the most elementary
level, i.~e.~by analysing the expansions of a simple Green's function such as
the nucleon self-energy. Such a 
comparison is hardly meaningful, however, since in general 
the Green's functions are model- and 
representation-dependent in perturbative as well as nonperturbative
approaches (see, e.g.,~\cite{Fea00} and references therein).
Instead, the values of the near-threshold parameters assembled  
in \tabref{chpt_comp} can be used to
illustrate the most important similarities and differences 
between the four approaches: 
the present model, HB$\chi$PT, RB$\chi$PT and the BSE.   

The basic dynamical degrees of freedom in all these approaches are the nucleon 
and the pion. However, unlike the two chiral formalisms, the present model
and the BSE include the $\Delta$ resonance and the $\rho$ and $\sigma$ mesons 
explicitly. In general, these degrees of freedom give important
contributions to the dressed Green's functions. In chiral approaches
effects of the resonances are typically encapsulated in low-energy
constants using a resonance-saturation hypothesis~\cite{Eck89,Ber95} 
(as mentioned above, we do not touch upon extensions~\cite{Ber93} of the chiral 
expansion with the $\Delta$ as an explicit degree of
freedom). In our model, as well as in
$\chi$PT, the $\pi N N$ vertex is predominantly pseudovector 
at low energies. 
A small pseudoscalar component in the
vertex is associated with an explicit symmetry breaking. In all four
approaches, the nontrivial dependence of the amplitudes
on kinematical variables is determined by loop expansions. 
The methods of organising the expansions are different however, as are
the methods of evaluating the loop diagrams. 

The chiral expansions are series in small pion mass and momenta~\cite{Wei79}. 
In the presence of a nucleon, a one-to-one correspondence
is established in HB$\chi$PT 
between the power of each term in the expansion and the number of pion 
loops required for its calculation~\cite{Jen91}. 
This nonrelativistic expansion is well-behaved 
in the part of the low-energy region which is not too close to the 
singularities of the amplitude dictated by unitarity and analyticity. 
However, it is
necessary to rearrange the chiral expansion in order to calculate the
amplitude near the singularities. This is most efficiently 
done in RB$\chi$PT~\cite{Bec01} by using an infrared regularisation in 
the relativistic formalism and representing the amplitude
through dispersion integrals. (Within HB$\chi$PT, the correct 
singularity structure can be captured only by summing an infinite number
of loop corrections.) The main emphasis of the approach of the BSE is on
obeying unitarity in the relativistic formalism~\cite{Sal51,Lah99}. 
The necessary loops
are summed up effectively by solving the linear integral equation
for the scattering amplitude. However, due to the use of a
simplified kernel in practical calculations, 
the BSE does not yield a crossing symmetric amplitude. 

The unitarity, crossing and analyticity constraints are used
as the governing principles for organising the meson-loop expansion in 
this model. The (nonlinear) dressing equations represent 
a self-consistent procedure
of using the cutting rules and dispersion relations. By solving them 
we effectively sum up an infinite series of loop diagrams in such a way that 
the essential singularity structure of the two- and three-point 
Green's functions is correctly reproduced. 
However, in contrast to the chiral
expansions or the BSE, one-particle irreducible four-point loop 
diagrams (such as the triangle or box graphs discussed in \cite{Bec01}) 
are not calculated explicitly in our model, which entails a violation of
analyticity structure in the t-channel. Also, in this model
we need to compute the 
dressed vertices only in the s- and u-type diagrams, 
i.~e.~for on-shell external pions only. Consequently,
the dependence of the amplitude on the momenta of external pions 
is not obtained. For instance, to obey
the Adler consistency condition~\cite{Adl68} 
$\overline{D}^+(\nu=0, t=\mu^2)=0$ (with either
$k^2=\mu^2, k^{\prime 2}=0$ or $k^2=0, k^{\prime 2}=\mu^2$),
the $\sigma N N$ vertex and the box graph 
will have to be dressed in this model,
thus requiring an extension to order ${\mathcal{O}}(a^2)$ in the
``analyticity violation expansion" outlined in \secref{anal}.    

In the calculation of RB$\chi$PT~\cite{Bec01}
the bulk of the sigma-term at the Cheng-Dashen point comes
from a contact four-point vertex. This contact vertex 
is proportional to the
pion mass squared and thus explicitly breaks chiral symmetry
of the lagrangian. In the present approach, 
the largest contribution to the sigma-term
is due to the t-channel $\sigma$ exchange with the
component $\sim g_{\sigma \pi \pi} \mu$ in the $\sigma \pi \pi$ 
vertex \eqref{sigpp}. This non-derivative
part of the $\sigma$ exchange does not vanish in the chiral limit and 
is therefore analogous to
the symmetry breaking contact term of RB$\chi$PT.
It should not be concluded, however, that
the loop contributions are of minor importance in the 
near-threshold region: it is through the loops that the essential unitarity, 
analyticity and crossing constraints are incorporated
in this model as well as in RB$\chi$PT.

Although the means of regularisation of the loop integrals
employed in this model and in the BSE (usage of bare form factors) 
are different to those utilised in the RB$\chi$PT 
(infrared and dimensional regularisation), they are
similar in that they generate spurious singularities of the amplitude.
These unwanted singularities 
should be removed from the region of physical interest. For example, the
amplitude in the RB$\chi$PT has an unphysical pole at $s=0$, 
which should be safely far from 
the relevant near-threshold region~\cite{Bec01}. 
Similarly, the bare form factor in this model is rather wide,
ensuring remoteness of its singularities.
There is an important difference between the evaluation
of the loop integrals in our approach and in the BSE. 
In the latter method the loops are
computed using such techniques as, e.~g.,~the Wick's rotation and
Feynman parametrisation, whereas in our framework the loops are calculated 
through the successive application of the Cutkosky rules and dispersion
relations. While being equivalent in local field theories,
these two methods of loop evaluation are likely to differ 
when the loops are regularised by form factors. Therefore  
the analytical properties of the amplitudes dressed in the BSE
are probably different to those generated by our approach,
which may have important consequences in the near-threshold region.

\section{Conclusions} \seclab{concl}

A consistent dynamical calculation of the pion-nucleon
amplitude in the near-threshold region should serve as a bridge between the
physically accessible low-energy data and chiral low-energy theorems 
reflecting the QCD dynamics in the nonperturbative regime.
In this paper we have shown that essential analyticity constraints
can be incorporated in a self-consistent dressing procedure,
resulting in a reliable description of the amplitude at the Cheng-Dashen point 
and at threshold. In particular, the pion-nucleon scattering lengths,
the nucleon sigma-term and the Adler-Weisberger coefficient $C$ 
evaluated in our approach are all consistent with the recent data analyses.

The values of the near-threshold coefficients depend crucially on 
the treatment of the $\sigma$ and $\rho$ meson exchanges.
The approach also includes
a consistent dressing of the $\pi N \Delta$ vertex and of 
the $\Delta$ propagator. This allows us to study the role of 
the $\Delta$ for the nucleon sigma-term in more detail than
has been done previously. In particular, we have found 
that the contribution of the $\Delta$ in the near-threshold region  
should be considered on a par with the 
effects of multiple loops. 

This dynamical model suggests that 
effective approaches 
incorporating the constraints of relativistic invariance, unitarity, 
crossing symmetry and analyticity can reveal important aspects of the 
low-energy strong interaction.

\acknowledgments
I would like to thank Harold Fearing, Andrew Lahiff, Olaf Scholten
and Marcello Pavan for stimulating discussions and helpful comments.
This work was supported in part by a grant from the Natural 
Sciences and Engineering Research Council of Canada.

\begin{appendix}

\section{Contributions of the $\rho$ and $\sigma$ to the dressing equations}
\seclab{contr_rhosig}

The term $\Gamma_{I}^{\rho \sigma}(p)$ on the right-hand side
of \eqref{dr_eq_n1} comprises 
6 loop diagrams including the $\rho$ and $\sigma$ degrees of 
freedom\footnote{In this appendix we use
the notation introduced in \secref{model}.},
\beq
{\ds \Gamma_{I}^{\rho \sigma}(p) = \sum_{n=1}^{3}\,
\Big\{\, (\Gamma_{I}^{\rho})_n(p) + (\Gamma_{I}^{\sigma})_n(p) \,\Big\} },
\eqlab{nucdr_rhosig}
\eeq
where the different terms are described in the following.
\beq
(\Gamma_{I}^{\rho})_1(p)=\frac{1}{8 \pi^2} 
\int d^4 k\, \Gamma^{\rho N N}_\mu(k-q,p-k) S_I(p-k)
\Gamma_{R}(p) D_\rho^{\mu \nu}(q-k) 
\Gamma^{\rho \pi \pi}_\nu(-k,q) D_I(k),
\eqlab{nucdr_rho1}
\eeq
where the $\rho \pi \pi$ vertex has the structure
\beq
\Gamma^{\rho \pi \pi}_\nu(q,q')=g_{\rho \pi \pi}
\left[q'_\nu-q_\nu + \frac{(q^2-q^{\prime 2})(q'_\nu+q_\nu)}{(q+q')^2}\right]
F_{\rho \pi \pi}((q+q')^2).
\eqlab{rhopp}
\eeq
In order that a converging solution of the dressing should exist, the
$\rho \pi \pi$ vertex is equipped with a form factor
$F_{\rho \pi \pi}((q+q')^2)$ whose form is the same as that of 
the bare $\pi N N$ form factor in \eqref{bareff},
\beq
F_{\rho \pi \pi}(k^2)=
\exp{\left[-\ln{2}\frac{(k^2-m_\rho^2)^2}{\Lambda_R^4}\right]},
\eqlab{mes_ff_bare}
\eeq
where the half-width $\Lambda_R^2$ is 
given in \tabref{param} and discussed in \secref{dres}.
By analogy with the treatment of the $\Delta$ 
resonance (see \eqref{gauge_inv}), 
we choose a gauge-invariant form of the $\rho \pi \pi$ vertex
and therefore retain only the spin 1 part of the $\rho$
propagator
\beq
D_\rho^{\mu \nu}(k)=\frac{{\mathcal{P}}_1^{\mu \nu}(k)}
{Z^\rho \,[\, k^2-\lambda^2(k^2) \,]},
\eqlab{rhoprop}
\eeq
with the spin 1 projector 
\beq
{\mathcal{P}}_1^{\mu \nu}(k)=g^{\mu \nu} - \frac{k^\mu k^\nu}{k^2}.
\eqlab{proj1}
\eeq
The $\rho$ self-energy function $\lambda^2(k^2)$ is calculated from 
one $\pi \pi$ loop as follows.
\beq
\lambda^2(p^2)=m_\rho^2 - \delta m_\rho^2 + 
\frac{\mbox{Re}\Pi^\rho_L(p^2)}{Z^\rho},
\eqlab{rho_lam}
\eeq
where $\Pi^\rho_L(p^2)$ is the $\pi \pi$ loop contribution 
to the self-energy and
$Z^\rho$, $\delta m_\rho^2$ are renormalisation constants
adjusted to ensure the correct pole properties of 
the dressed $\rho$ propagator \eqref{rhoprop}.  
As all loop integrals in the model, this loop is evaluated
using a dispersion relation:  
\beq
\mbox{Im}\Pi^{\rho}(k^2)=-\frac{{\mathcal{P}}_1^{\nu \mu}(k)}{24 \pi^2} 
\int d^4 q \, \Gamma^{\rho \pi \pi}_\mu(q,k-q) D_I(q) D_I(k-q) 
\Gamma^{\rho \pi \pi}_\nu(q-k,-q) \,,
\eqlab{rho_im}
\eeq
\beq
\mbox{Re}\Pi^{\rho}_L(k^2)=\frac{\mathcal{P}}{\pi}
\int_{4 \mu^2}^{\infty} \!\! dk^{\prime 2}\,
\frac{\mbox{Im} \Pi^\rho(k^{\prime 2})}{k^{\prime 2}-k^2}\;.
\eqlab{rho_re}
\eeq
The $\rho N N$ vertex is chosen as
\beq
\Gamma^{\rho N N}_\mu(k,p)=g_{\rho N N}
\left[ \gamma_\mu + i \kappa_\rho \frac{\sigma_{\mu \eta}k^\eta}{2 m} \right]
F_{\rho N N}(p^2),
\eqlab{rhonn}
\eeq
where the regularising form factor $F_{\rho N N}(p^2)$ depends on the
four-momentum squared of an off-shell nucleon and has the form of
\eqref{mes_ff_bare} in which $m_N$ is substituted for $m_\rho$. 
Describing \eqref{nucdr_rhosig} further,
\beq
(\Gamma_{I}^{\rho})_2(p)=\frac{g \gamma_5}{8 \pi^3} 
\int d^4 k\, S_I(p-k) \Gamma^{\rho N N}_\mu(-k,p) 
D(k-q) \Gamma^{\rho \pi \pi}_\nu(-q,q-k)  (D_\rho^{\mu \nu})_I(k),
\eqlab{nucdr_rho2}
\eeq
where \eqref{vert_renorm} has been used;
\beq
(\Gamma_{I}^{\rho})_3(p)=
\frac{1}{32 \pi^3} \int d^4 k\, 
\Gamma^{\rho N N}_\mu(k,p'-k) S(p'-k) \overline{\Gamma}_{R}(p'-k) S_I(p-k)  
\Gamma^{\rho N N}_\nu(-k,p) 
(D_\rho^{\mu \nu})_I(k) \,.
\eqlab{nucdr_rho3}
\eeq
The loop integrals with the $\sigma$ meson are similar in structure to those with the $\rho$.
\beq
(\Gamma_{I}^{\sigma})_1(p)=-\frac{g_{\sigma N N}}{8 \pi^2} 
\int d^4 k\, S_I(p-k) \Gamma_{R}(p) D_\sigma(q-k)  
\Gamma^{\sigma \pi \pi}(-k,q) D_I(k),
\eqlab{nucdr_sig1}
\eeq
where the $\sigma \pi \pi$ vertex is chosen as
\beq
\Gamma^{\sigma \pi \pi}(q,q')= \left[ g_{\sigma \pi \pi} \mu + 
f_{\sigma \pi \pi} \frac{q \cdot q'}{\mu} \right] F_{\sigma \pi \pi}((q+q')^2).
\eqlab{sigpp}
\eeq 
The $\sigma$ propagator $D_\sigma(k)$ is obtained from one 
$\pi \pi$ loop:
\beq
D_\sigma(k)=\frac{1}{Z^\sigma \, [\, k^2-\zeta^2(k^2) \,]},
\eqlab{sigmaprop}
\eeq  
\beq
\zeta^2(p^2)=m_\sigma^2 - \delta m_\sigma^2 + 
\frac{\mbox{Re} \Pi_L^\sigma(p^2)}{Z^\sigma},
\eqlab{sigma_zeta}
\eeq
\beq
\mbox{Im}\Pi^{\sigma}(k^2)=-\frac{3}{16 \pi^2} 
\int d^4 q \, \Gamma^{\sigma \pi \pi}_\mu(q,k-q) D_I(q) D_I(k-q) 
\Gamma^{\sigma \pi \pi}_\nu(q-k,-q) \,,
\eqlab{sigma_im}
\eeq
\beq
\mbox{Re}\Pi^{\sigma}_L(k^2)=\frac{\mathcal{P}}{\pi}
\int_{4 \mu^2}^{\infty} \!\! dk^{\prime 2}\,
\frac{\mbox{Im} \Pi^\sigma(k^{\prime 2})}{k^{\prime 2}-k^2}\;.
\eqlab{sigma_re}
\eeq
The two other loops with the $\sigma$ read
\beq
(\Gamma_{I}^{\sigma})_2(p)=
\frac{g \gamma_5}{8 \pi^3} 
\int d^4 k\, S_I(p-k) \Gamma^{\sigma N N}(-k,p) D(k-q) 
\Gamma^{\sigma \pi \pi}(-q,q-k) (D_\sigma)_I(k),
\eqlab{nucdr_sig2}
\eeq
\beq
(\Gamma_{I}^{\sigma})_3(p)=
\frac{g_{\sigma N N}^2 }{8 \pi^3} \int d^4 k\,  
\Gamma^{\sigma N N}(k,p'-k) S(p'-k) \overline{\Gamma}_{R}(p'-k) S_I(p-k) 
\Gamma^{\sigma N N}(-k,p) (D_\sigma)_I(k),
\eqlab{nucdr_sig3}
\eeq
where the $\sigma N N$ vertex has the simple structure 
\beq
\Gamma^{\sigma N N}(k,p)=g_{\sigma N N} F_{\sigma N N}(p^2),
\eqlab{signn}
\eeq
with the form factor $F_{\sigma N N}(p^2)=F_{\rho N N}(p^2)$.

The term $(V^{\rho \sigma}_\mu)_I(q,p)$ in \eqref{dr_eq_d1}
comprises two loop integrals whose
form is given by \eqref{nucdr_rho1} and \eqref{nucdr_sig1} 
in which the $\pi N N$ vertex $\Gamma_R(p)$ is
replaced with the $\pi N \Delta$ vertex $(V_\mu)_R(k,p)$.

\section{Contribution of the $\Delta$ at the Cheng-Dashen point}
\seclab{contr_delta}

The contributions of the $\Delta$ resonance to the
sigma-term and to the coefficient $C$ in 
Eqs.~(\ref{eq:sigterm_calc},\ref{eq:coefc_calc}) 
are obtained by decomposing Eqs.~(\ref{eq:dels},\ref{eq:delu}) into
the invariant amplitudes $D^{\pm}$
and evaluating the latter at the Cheng-Dashen point. 
The expressions in terms of the dressed $\pi N \Delta$ form factor
and the dressed $\Delta$ self-energy functions (as defined in
Eqs.~(\ref{eq:delvert},\ref{eq:delprop})) are 
\begin{eqnarray}
\Sigma_\Delta&=&-\frac{2 F_\pi^2 G_\Delta^2(m^2) \, \omega(m^2)}
{9 \, m_\Delta^4 \eta(m^2) [\, m^2-\omega^2(m^2) \,]} \, \mu^4 \,,
\eqlab{sig_del_expl} \\
C_\Delta&=&\frac{8 F_\pi^2 m \omega(m^2) G_\Delta^2(m^2)}
{9\, m_\Delta^4 \eta(m^2) [\, m^2-\omega^2(m^2)  \,]}\, \mu^2  \nn \\
&+&2 F_\pi^2 
\left\{ \frac{G_\Delta^2(m^2)+4 m \omega(m^2) G_\Delta(m^2) G_\Delta'(m^2)}
{9 m_\Delta^4 \eta(m^2) [\, m^2-\omega^2(m^2) \,]} 
- \frac{2 m G_\Delta^2(m^2) \eta(m^2) \omega(m^2) }
{9 m_\Delta^4 \eta^2(m^2) [\, m^2-\omega^2(m^2) \,]^2} \right.  \nn  \\
&+& \left. \frac{2 m G_\Delta^2(m^2) [\, m^2+\omega^2(m^2) \,] \,
[\, \eta(m^2) \omega'(m^2) - \omega(m^2) \eta'(m^2) \,] }
{9 m_\Delta^4 \eta^2(m^2) [\, m^2-\omega^2(m^2) \,]^2} \right\} \, \mu^4 \,,
\eqlab{sig_c_expl}
\end{eqnarray}
explicitly showing the suppression by powers of the pion mass $\mu$.

\end{appendix}


\begin{table}
\caption[t1]{Particle masses (in GeV) and coupling constants used in this 
calculation (the same values were used in Ref.~\cite{Kon01}). 
The masses and the coupling constants of the $\Delta$, 
$\rho$ and $\sigma$ correspond to the experimental widths and positions of the
resonances as given in \cite{PDG00}.}
\begin{center}
\begin{tabular}{cccccccc}
\hline\hline
\hspace*{3mm}$m \equiv m_N$ \hspace*{3mm} & \hspace*{3mm} $\mu \equiv m_\pi$ 
\hspace*{3mm} 
&\hspace*{3mm} $m_\Delta$ \hspace*{3mm} &\hspace*{3mm} $m_\rho$ \hspace*{3mm} 
&\hspace*{3mm} $m_\sigma$ \hspace*{3mm} & \hspace*{3mm} $g \equiv g_{\pi N N}$ 
\hspace*{3mm} &\hspace*{3mm} $g_{\pi N \Delta}$ \hspace*{3mm} &\hspace*{3mm} $g_{\rho \pi \pi}$ \hspace*{3mm} \\
\hline
 $0.939$ & $0.138$  & $1.232$ & $0.77$ & $0.76$ & $13.02$ & $19.76$ & $6.07$ \\
\hline\hline
\end{tabular}
\end{center}
\tablab{masscoupl}
\end{table}

\begin{table}
\caption[t2]{The renormalisation parameters:
bare coupling constants, mass shifts (in units of GeV) 
and field renormalisation factors.}
\begin{center}
\begin{tabular}{cccccccccc}
\hline\hline
\hspace*{1mm} $f \equiv f_{\pi N N}$ \hspace*{1mm} 
& \hspace*{1mm} $f_{\pi N \Delta}$ \hspace*{1mm} 
&\hspace*{1mm} $\delta m \equiv \delta m_N$ \hspace*{1mm} 
&\hspace*{1mm} $\delta m_\Delta$ \hspace*{1mm} 
&\hspace*{1mm} $\delta m_\rho^2$ \hspace*{1mm}
&\hspace*{1mm} $\delta m_\sigma^2$ \hspace*{1mm} 
& \hspace*{1mm} $Z_2 \equiv Z_2^N$ \hspace*{1mm}
&\hspace*{1mm} $Z_2^\Delta$ \hspace*{1mm} &\hspace*{1mm} $Z^\rho$ \hspace*{1mm} 
&\hspace*{1mm} $Z^\sigma$ \hspace*{1mm} \\
\hline
$10.75$ & $21.75$ & $-0.075$ & $-0.120$ & $-0.089$ & $-0.605$ & $0.80$ 
& $1.16$ & $1.17$ & $1.14$ \\
\hline\hline
\end{tabular}
\end{center}
\tablab{renorm}
\end{table}

\begin{table}
\caption[t3]{Values of the parameters of the model, as fixed
by calculating the intermediate-energy pion-nucleon phase shifts
in the dressed K-matrix approach of Ref.~\cite{Kon01}.
No parameters were readjusted in the present calculation.}
\begin{center}
\begin{tabular}{ccccccc}
\hline\hline
\hspace*{3mm} $\Lambda^2_N$ \hspace*{3mm} 
&\hspace*{3mm} $\Lambda^2_R$ \hspace*{3mm}
&\hspace*{3mm} $g_{\rho N N}$ \hspace*{3mm}  
&\hspace*{3mm} $\kappa_\rho$ \hspace*{3mm}   
&\hspace*{3mm} $g_{\sigma N N}$ \hspace*{3mm} 
&\hspace*{3mm} $g_{\sigma \pi \pi}$ \hspace*{3mm} 
&\hspace*{3mm} $f_{\sigma \pi \pi}$ \hspace*{3mm} \\
\hline
$1.8$  & $1.0$ & $7.0$ & $2.3$ & $34$ & $1.7$ & $1.8$   \\
\hline\hline
\end{tabular}
\end{center}
\tablab{param}
\end{table}

\begin{table}
\caption[t4_mod]{The pion-nucleon sigma-term, the Adler-Weisberger coefficient $C$, 
evaluated at the Cheng-Dashen point 
from Eqs.~(\ref{eq:sigterm},\ref{eq:coefc}),
and the s-wave scattering lengths, evaluated from 
Eqs.~(\ref{eq:a1},\ref{eq:a3}). The rows represent the following calculations. 
``Dressed": fully dressed calculation; 
``Bare": bare calculation, i.~e.~using the free propagators and 
no loop corrections to the bare vertices;
``Free $\sigma$": full calculation, but using the free $\sigma$ propagator in the t-channel exchange;
``Free $\rho$": full calculation, but using the free $\rho$ propagator in the t-channel;
``No $\Delta$ poles": full calculation, but without
the s- and u-channel $\Delta$ exchange pole diagrams;
``Bare $\Delta$": full calculation, but using the bare $\pi N \Delta$ vertex and the free
$\Delta$ propagator;
``One loop": calculation in which the nucleon and $\Delta$ self-energies as well as the
$\pi N N$ and $\pi N \Delta$ vertices are computed up to one-loop corrections only;
``Data": results of various data analyses.} 
\begin{center}
\begin{tabular}{l c c c c}
\hline\hline
        & $\Sigma$ (MeV) & $C$    & $a^{1/2}(\mu^{-1})$ & $a^{3/2}(\mu^{-1})$ \\
\hline
Dressed & $73.99$        & $1.16$ & $0.175$             & $-0.087$            \\
\hline
Bare    & $127.78$       & $1.31$ & $0.204$             & $-0.088$            \\
\hline
Free $\sigma$ & $126.41$ & $1.16$ & $0.183$             & $-0.080$            \\
\hline
Free $\rho$   & $73.99$  & $1.40$ & $0.210$             & $-0.105$            \\
\hline
No $\Delta$ poles & $73.73$ & $1.21$ & $0.175$          & $-0.087$            \\
\hline
Bare $\Delta$ & $74.05$ & $1.14$ & $0.175$              & $-0.087$            \\
\hline
One loop      & $71.98$ & $1.21$ & $0.181$              & $-0.093$            \\
\hline
Data          & \hspace*{4mm}$64 \pm 8$ \cite{Hoh83} & \hspace*{4mm}$1.15 \pm 0.02$ \cite{Hoh83} 
& \hspace*{3mm} $0.173 \pm 0.003$ \cite{Eri88} & \hspace*{2mm} $-0.101 \pm 0.004$ \cite{Eri88} \\  
              & \hspace*{4mm}$79 \pm 7$ \cite{Pav01} & & \hspace*{6.5mm}$0.175 \pm 0.004$ \cite{Sch99} 
	                                   & \hspace*{4mm}$-0.085 \pm 0.027$ \cite{Sch99}  \\
              & \hspace*{2mm}$71 \pm 9$ \cite{Ols00}	& & &                                 \\
\hline
\hline	      
\end{tabular}
\end{center}
\tablab{thresh_mod}
\end{table}

\begin{table}
\caption[t5]{Comparison of the near-threshold parameters
evaluated in the present model with 
results obtained in chiral perturbation theory and in the approach based
on the Bethe-Salpeter equation.
The third and fourth order HB$\chi$PT calculations~\cite{Fet98} 
presented fits to three different phase-shifts, using 
Ref.~\cite{Ols00} to relate the sigma-term and threshold parameters. 
The RB$\chi$PT calculation~\cite{Bec01} used
the data analyses of~\cite{Hoh83} as input. The 
BSE results are from Ref.~\cite{Lah99}.}
\begin{center}
\begin{tabular}{l c c c c c}
\hline\hline
  \hspace*{3mm} & This model \hspace*{3mm} 
  & HB$\chi$PT ${\mathcal{O}}(p^3)$ \hspace*{3mm}
  & HB$\chi$PT ${\mathcal{O}}(p^4)$ \hspace*{3mm}  
  & RB$\chi$PT ${\mathcal{O}}(p^4)$ \hspace*{3mm} & BSE \\
\hline
$\Sigma$ (MeV)   & $73.99$  & $69$ 
& $73$  & $61$ & $23.6$  \\
 & & $91$ & $85$ & &       \\
 & & $93$ & $104$ & &      \\ 
\hline 
$C$                   & $1.16$     & $1.10$  &      & $1.13$ &    \\ 
 & & $0.82$ & & & \\
 & & $1.09$ & & & \\ 
\hline
$a^{1/2}(\mu^{-1})$ & $0.175$    & $0.171$  
& $0.171$       & $0.175$ & $0.177$  \\
 & & $0.159$ & $0.159$ &  & \\
 & & $0.175$ & $0.176$ &  & \\
\hline
$a^{3/2}(\mu^{-1})$ & $-0.087$   & $-0.101$
& $-0.100$       & $-0.100$  & $-0.101$ \\
 & & $-0.072$ & $-0.073$ & & \\
 & & $-0.086$ & $-0.084$ & & \\
\hline\hline
\end{tabular}
\end{center}
\tablab{chpt_comp}
\end{table}

\Omit{  
\begin{table}
\caption[t1]{Nucleon sigma-term, the low-energy quantity $C$, 
calculated at the Cheng-Dashen point 
from Eqs.~(\ref{eq:sigterm}) and (\ref{eq:coefc}),
and the s-wave scattering lengths, calculated from the
Eqs.~(\ref{eq:a1}) and (\ref{eq:a3}). The following five calculations are
presented. Column 1: the fully dressed calculation; column 2:
the full calculation without the consistent dressing of the $\Delta$ propagator
and $\pi N \Delta$ vertex; column 3: the full calculation without
the s- and u-channel $\Delta$ exchange pole diagrams; column 4: 
the calculation in which the nucleon, $\Delta$ self-energies and the
$\pi N N$, $\pi N \Delta$ vertices are
computed up to one-loop corrections only; column 5: bare calculation, i.e.\
using the free
propagators and no loop corrections to the bare vertices.} 
\begin{center}
\begin{tabular}{l c c c c c c}
\hline\hline
\hspace*{2mm} & Dressed \hspace*{2mm} & Bare $\Delta$ \hspace*{2mm} 
& No $\Delta$ poles \hspace*{2mm} &  One loop \hspace*{2mm}
&  Bare \hspace*{2mm} & Data analyses \\
\hline
$\Sigma$ (MeV) &  $63.94$ & $64.00$ & $63.68$ & $69.74$ 
&  $109.63$  & $64 \pm 8$ \cite{Hoh83}      \\
 &  &  &  &  &  & $71 \pm 9$ \cite{Ols00}     \\
 &  &  &  &  &  & $79 \pm 7$ \cite{Pav01}   \\
\hline 
$C$  &  $1.16$  &  $1.15$  & $1.21$ & $1.40$ 
&  $1.31$  & $1.15 \pm 0.02$ \cite{Hoh83}   \\
\hline
$a^{1/2}(\mu^{-1})$ & $0.168$ & $0.168$ & $0.168$ & $0.202$
&   $0.192$   & $\;\;\, 0.173 \pm 0.0003$ \cite{Eri88}     \\
& & & & & & $ \;\;\, 0.175 \pm 0.0041$ \cite{Sch99} \\
\hline
$a^{3/2}(\mu^{-1})$ & $-0.095$ & $-0.095$ & $-0.095$ & $-0.114$
&  $-0.100$ & $\;\;\, -0.101 \pm 0.0004$ \cite{Eri88} \\
& & & & & & $ \;\;\, -0.085 \pm 0.0027$ \cite{Sch99} \\
\hline\hline
\end{tabular}
\end{center}
\tablab{thresh}
\end{table}

\begin{table}
\caption[t2]{Comparison of the near-threshold parameters
calculated in the present model with 
results obtained in the chiral perturbation theory.
The heavy-baryon calculation (HB$\chi$PT) of~\cite{Fet98}
presents three fits to phase-shifts. The relativistic infrared-regularised 
approach (RB$\chi$PT) is from~\cite{Bec01}.}
\begin{center}
\begin{tabular}{l c c c c}
\hline\hline
  \hspace*{3mm} & This model \hspace*{3mm} 
  & HB$\chi$PT ${\mathcal{O}}(p^3)$ \hspace*{3mm}
  & HB$\chi$PT ${\mathcal{O}}(p^4)$ \hspace*{3mm}  
  & RB$\chi$PT ${\mathcal{O}}(p^4)$ \\
\hline
$\Sigma$ (MeV)   & $63.94$  & $62$(fit 1) 
& $73$(fit 1)  & $61$   \\
 & & $79$(fit 2) & $85$(fit 2) &        \\
 & & $82$(fit 3) & $104$(fit 3) &       \\ 
\hline 
$C$                   & $1.16$     & $1.10$(fit 1)  &      & $1.13$    \\ 
 & & $0.82$(fit 2) & & \\
 & & $1.09$(fit 3) & & \\ 
\hline
$a^{1/2}(\mu^{-1})$ & $0.168$    & $0.171$(fit 1)  
& $0.171$(fit 1)       & $0.175$   \\
 & & $0.159$(fit 2) & $0.159$(fit 2) &  \\
 & & $0.175$(fit 3) & $0.176$(fit 3) &  \\
\hline
$a^{3/2}(\mu^{-1})$ & $-0.095$   & $-0.101$(fit 1)
& $-0.100$(fit 1)       & $-0.100$  \\
 & & $-0.072$(fit 2) & $-0.073$(fit 2) & \\
 & & $-0.086$(fit 3) & $-0.084$(fit 3) & \\
\hline\hline
\end{tabular}
\end{center}
\tablab{chpt_comp}
\end{table}
}


\begin{figure}
\centerline{{\epsfxsize 15cm \epsffile[50 380 565 450]{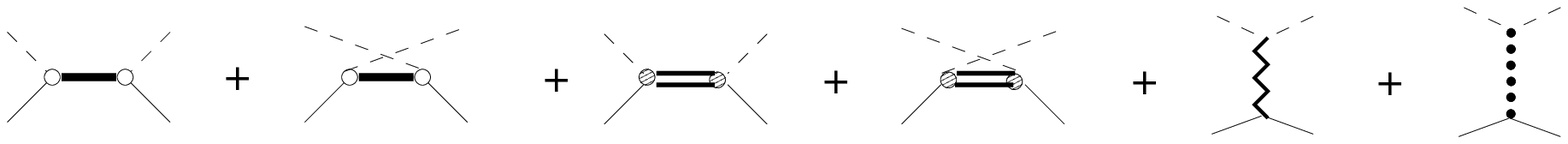}}}
\caption[f1]{Graphical representation of the
pion-nucleon amplitude, corresponding to \eqref{amp_sut}. 
The single solid lines are nucleons, the double lines 
are $\Delta$'s,
the dashed, zigzag and dotted lines are pions, $\rho$'s and $\sigma$'s,
respectively. The empty and hatched circles denote the 
$\pi N N$ and $\pi N \Delta$ vertices, respectively.
The propagators and vertices are dressed with meson loops as
described in \secref{dres}.
\figlab{pin_pnn_pnd}}
\end{figure}

\newpage

\begin{figure}
\centerline{{\epsfxsize 15cm \epsffile[5 80 570 735]{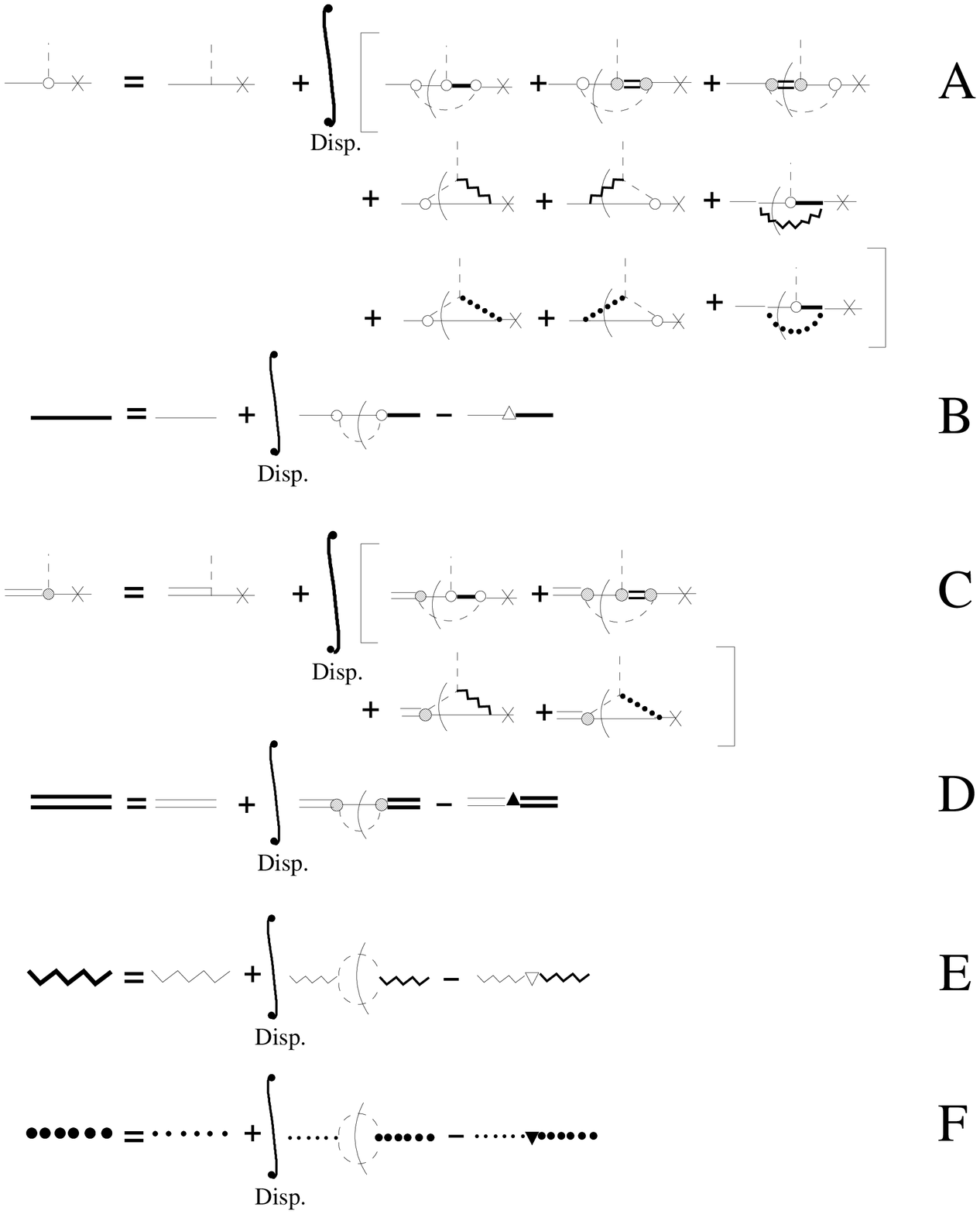}}}
\caption[f2]{Graphical representation of the system of integral 
equations for the dressed two- and three-point Green's functions.
The notation for the propagators and vertices is as in \figref{pin_pnn_pnd}. 
For each particle, the dressed and free propagators are denoted by thick and
thin lines, respectively. 
The triangles denote
the counterterms needed to fulfill the renormalisation
conditions such as \eqref{prop_renorm}.
The slashes through the loops and the integral signs indicate the 
use of the Cutkosky (cutting) rules and dispersion integrals in the 
iterative solution of the equations.
The outgoing nucleons in the vertices (as well as the pions) are on-shell,
which is denoted by the crossed lines.  
The correspondence with analytic equations is 
$\mbox{\bf A} \leftrightarrow (\ref{eq:dr_eq_n1},\ref{eq:dr_eq_n2})$,
$\mbox{\bf B} \leftrightarrow (\ref{eq:dr_eq_n3},\ref{eq:dr_eq_n4})$,    
$\mbox{\bf C} \leftrightarrow (\ref{eq:dr_eq_d1},\ref{eq:dr_eq_d2})$,
$\mbox{\bf D} \leftrightarrow (\ref{eq:dr_eq_d3},\ref{eq:dr_eq_d4})$,
$\mbox{\bf E} \leftrightarrow (\ref{eq:rho_im},\ref{eq:rho_re})$,
$\mbox{\bf F} \leftrightarrow (\ref{eq:sigma_im},\ref{eq:sigma_re})$.
\figlab{eq_pnn_pnd}}
\end{figure}

\begin{figure}
\centerline{{\epsfxsize 13.5cm \epsffile[35 230 570 620]{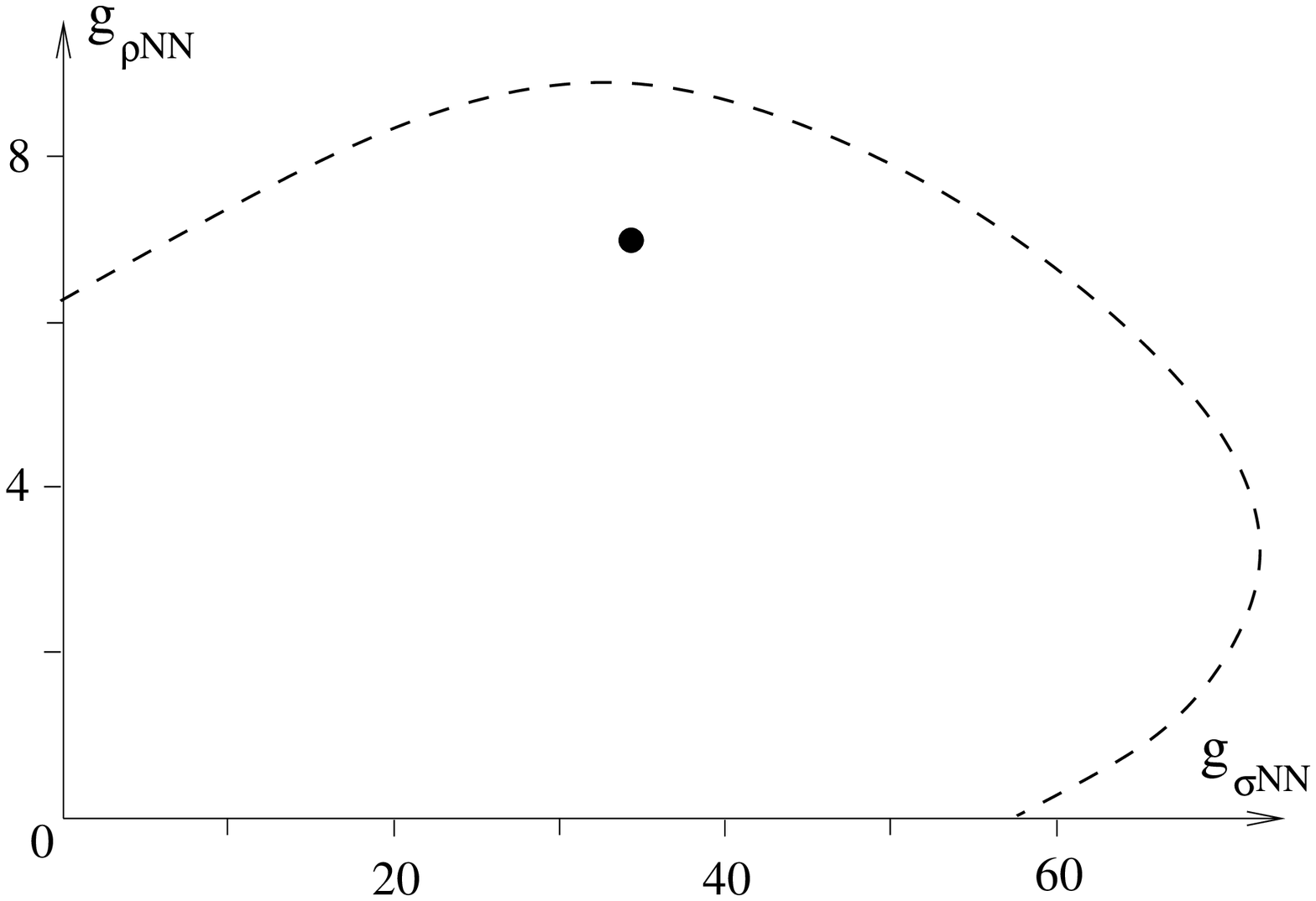}}}
\caption[f3]{Illustration of the interdependence of the coupling 
constants $g_{\sigma N N}$ and $g_{\rho N N}$ 
(the other parameters being fixed as in \tabref{param})
due to the requirement of convergence of the dressing procedure.
The area of convergence is sketched using 50 test solutions of
Eqs.~(\ref{eq:dr_eq_n1}--\ref{eq:dr_eq_d4}), 
but ignoring $\pi N$ phase shifts. The dot corresponds to the
phase shifts fit obtained in Ref.~\cite{Kon01}. 
\figlab{param_dep}}
\end{figure}


\end{document}